\documentclass[11pt,a4paper]{article}
\pdfoutput=1

\usepackage{jheppub}
\usepackage{amsmath}

\graphicspath{ {fig/} }

\newcommand{\p}{p \hspace{-1.0ex} /}
\newcommand{\M}{ {\rm M} }
\newcommand{\e}{ {\rm e} }
\newcommand{\un}{ {\rm un} }
\newcommand{\finite}{ {\rm finite} }
\newcommand{\C}{ \mathcal{C} }
\newcommand{\F}{ \mathcal{F} }
\newcommand{\MS}{\overline{ \rm MS}}
\newcommand{\ud}{\mathrm{d}}
\newcommand{\mm}{\bar{m}}
\newcommand{\xx}{\bar{x}}
\newcommand{\yy}{\bar{y}}
\newcommand{\zz}{\bar{z}}
\newcommand{\reduze}{{\tt Reduze}}
\newcommand{\hepforge}{{\tt HepForge}}
\newcommand{\qgraf}{{\tt Qgraf}}
\newcommand{\openloops}{{\tt OpenLoops}}
\newcommand{\form}{{\tt Form}}
\newcommand{\jaxodraw}{{\tt JaxoDraw}}
\DeclareMathOperator{\Li}{Li}
\DeclareMathOperator{\G}{G}

\allowdisplaybreaks[3]

\makeatletter
\renewcommand\@fpheader{} 
\renewcommand\@journal{}
\makeatother

\title{
The two-loop helicity amplitudes for $q \bar{q}' \to V_1 V_2\to 4~\mathrm{leptons}$
}

\preprint{{ZU-TH 03/15, MITP/15-011, TTP15-011}}

\author[a]{Thomas Gehrmann,}
\author[b]{Andreas von Manteuffel,}
\author[c]{Lorenzo Tancredi}

\affiliation[a]{
  Physik-Institut, 
  Universit\"at Z\"urich, Wintherturerstrasse 190,
  CH-8057~Z\"urich, Switzerland}

  \affiliation[b]{
  PRISMA Cluster of Excellence, Institute of Physics,
  Johannes Gutenberg University,\\
  55099~Mainz, Germany}
  
  \affiliation[c]{
  Institut f\"{u}r Theoretische Teilchenphysik, Karlsruhe 
  Institute of Technology,\\ Engesserstrasse 7, 76128 Karlsruhe, Germany
  }

\emailAdd{thomas.gehrmann@uzh.ch}
\emailAdd{manteuffel@uni-mainz.de}
\emailAdd{lorenzo.tancredi@kit.edu}

\keywords{QCD, Collider Physics, NLO and NNLO Calculations}

\abstract{
We compute the two-loop massless QCD corrections to the helicity amplitudes for the production of two massive vector 
bosons in quark-antiquark annihilation, allowing for an arbitrary virtuality of the vector bosons:
$q\bar{q}' \to V_1V_2$. Combining with the leptonic 
decay currents, we obtain the full two-loop QCD description of the corresponding electroweak four-lepton production processes.
The calculation is performed by projecting the two-loop diagrams onto an appropriate basis of Lorentz structures. All two-loop 
Feynman integrals are reduced to a basis of master integrals, which are then computed using the differential equations method
and optimised for numerical performance.
We provide a public {\tt C++} code which allows for fast and precise numerical evaluations of the amplitudes.
}

\begin{document}
\unitlength1cm
\maketitle

\section{Introduction}
\label{sec:intro}

Vector boson pair production is an outstandingly important process at high energy hadron colliders. Its measurement allows 
precision studies of the electroweak interaction, thereby testing in detail the $SU(2)_L\times U(1)_Y$ gauge structure and the 
matter content of the Standard Model of particle physics. The various combinations of vector boson pairs 
($ZZ$, $W^+W^-$, $\gamma\gamma$, $ZW^\pm$, $Z\gamma$, $W^\pm\gamma$) lead to spectacular final state signatures 
(leptons, photons, missing energy),  that are often equally relevant to searches for new physics
or studies of the Higgs boson. The  
Higgs boson decay into two vector bosons is among the cleanest signatures for Higgs production, and offers a broad spectrum 
of observables. 

Precision studies of the electroweak interaction often  focus on the pair production of on-shell
gauge bosons, while new physics searches and Higgs boson studies precisely veto these on-shell contributions, such that 
the remaining background processes are dominated by off-shell gauge boson pair production. For both on-shell and 
off-shell production processes, it is therefore very important to have a precise prediction of the Standard Model 
contributions, in order to match the anticipated experimental accuracy of measurements at the LHC, which is usually in the 
per-cent range. At this level of precision, next-to-leading order (NLO) corrections in the electroweak theory and 
next-to-next-to-leading order (NNLO) corrections in QCD are indispensable. 

For all vector boson pair production processes, NLO QCD 
corrections~\cite{Ohnemus:1992jn,Baur:1993ir,Baur:1997kz,Dixon:1998py,Campbell:1999ah,Dixon:1999di}  as well as large parts of the 
NLO electroweak 
corrections~\cite{Accomando:2004de,Accomando:2005xp,Accomando:2005ra,Bierweiler:2013dja,Baglio:2013toa,Billoni:2013aba,Gieseke:2014gka,Denner:2014bna} 
are available. These calculations are fully differential in all kinematical variables, and usually include the leptonic 
decays of the vector bosons. 
The derivation of NNLO QCD corrections to vector boson pair production can build upon 
calculational techniques~\cite{Anastasiou:2005qj,Catani:2007vq}
that were originally developed for the Drell-Yan process~\cite{Melnikov:2006kv,Catani:2009sm} or for Higgs boson 
production in gluon fusion~\cite{Anastasiou:2005qj,Catani:2007vq}, which 
have the same QCD structure due to their colour-neutral final state.
As a new ingredient, each vector boson 
pair production process at NNLO requires the two-loop corrections to the basic scattering amplitude for 
quark-antiquark annihilation: $q\bar{q}'\to V_1V_2$. These have been known for a while already for 
$\gamma \gamma$~\cite{Bern:2001df,Anastasiou:2002zn} and 
$V\gamma$~\cite{Gehrmann:2011ab,Gehrmann:2013vga} production, enabling the 
calculations of these processes~\cite{Catani:2011qz,Grazzini:2013bna} to NNLO accuracy. 

Compared to the above, the two-loop matrix elements for the production of a pair 
of massive vector bosons require a 
new class of two-loop Feynman integrals: two-loop four-point functions with  
massless internal propagators and two 
massive external legs. Recently, very important progress has been made on these. 
For the case of equal vector 
boson mass, these integrals were derived in~\cite{Gehrmann:2013cxs,Gehrmann:2014bfa}, and used subsequently 
to compute the NNLO corrections to the on-shell production of $ZZ$~\cite{Cascioli:2014yka} 
and $W^+W^-$~\cite{Gehrmann:2014fva}. The integrals for the 
most general case of non-equal masses were derived in~\cite{Henn:2014lfa,Caola:2014lpa,Papadopoulos:2014hla},
which allowed to construct the full two-loop helicity amplitude for $q\bar{q}'\to V_1V_2$ in~\cite{Caola:2014iua}.
A subset of these integrals was derived independently in~\cite{Chavez:2012kn,Anastasiou:2014nha} and used in the derivation of the fermionic NNLO corrections to $\gamma^*\gamma^*$ production~\cite{Anastasiou:2014nha}.
In this paper, we  perform an independent rederivation of these integrals
and optimise our solutions for numerical performance.
They are used subsequently for a validation 
of  the two-loop helicity amplitudes of~\cite{Caola:2014iua}, uncovering an error in their original results. 
We present a public implementation for the numerical evaluation of
these amplitudes, which in the future will allow the calculation of NNLO QCD
corrections to arbitrary electroweak four-fermion production processes.

The paper is structured as follows: 
in Section~\ref{sec:amp}, we introduce the partonic current for
vector boson pair production and describe its decomposition into
Lorentz structures.
Taking into account the vector boson decays into leptons,
we present the helicity amplitudes for four particle final state
in Section~\ref{sec:hel}.
A detailed description of the calculation and the different
contributions to the amplitude is given in Section~\ref{sec:calc}.
The computation of the master integrals and their optimisation is presented in~\ref{sec:masters}.
In Section~\ref{sec:helfin}, we describe the subtraction of UV and IR
counter terms, and in Section~\ref{sec:checks} we list the numerous checks
we performed on our results.
In Section~\ref{sec:numerics} we present our {\tt C++} implementation
for the numerical evaluation of the amplitudes and use it to
produce numerical results.
Finally, we conclude in Section~\ref{sec:conc}.
In Appendix~\ref{sec:equalmass}, we document  the interference of the two-loop and tree amplitudes for the production
of on-shell vector boson pairs, which was used in the calculation of the NNLO corrections to 
$pp\to ZZ$~\cite{Cascioli:2014yka} and 
$pp\to WW$~\cite{Gehrmann:2014fva}. Appendix~\ref{sec:schouten} contains the derivation of 
Schouten identities for the leptonic
amplitudes, and Appendix~\ref{sec:catani} describes the conversion of our results between different schemes for the subtraction of infrared singularities. 
We provide computer readable files for our analytical results
and our {\tt C++} code for the numerical evaluation of the amplitude
on our {\tt VVamp} project page on \hepforge\ at \url{http://vvamp.hepforge.org}.

\section{\texorpdfstring
{Lorentz structure of the partonic current for $q\bar{q}'\to V_1 V_2$ }
{Lorentz structure of the partonic current for qq' -> V1 V2}
}
\label{sec:amp}

Let us consider the production of two massive electroweak vector bosons in $q \bar{q}'$ annihilation:
\begin{equation}
q(p_1) + \bar{q}'(p_2) \longrightarrow V_1(p_3) + V_2(p_4)
\end{equation}
with
\begin{align}
p_1^2 = p_2^2 = 0\,, \qquad p_3^2 \neq 0\,, \qquad p_4^2 \neq 0\,,
\end{align}
where the two vector bosons are off-shell and $V_1V_2$ = 
$ZZ$, $W^+W^-$, $\gamma\gamma$, $ZW^\pm$, $Z\gamma$, $W^\pm\gamma$.
We define the usual Mandelstam variables
\begin{equation}
s=(p_1+p_2)^2\,, \qquad t=(p_1-p_3)^2\,,\qquad u=(p_2-p_3)^2\,,
\end{equation}
such that $$s+t+u = p_3^2 + p_4^2\,.$$
The physical region of phase space is bounded by $t u = p_3^2 p_4^2$
such that
\begin{equation}
s \geq \Big(\sqrt{p_3^2} + \sqrt{p_4^2}\Big)^2,\qquad
\frac{1}{2}\big(p_3^2+p_4^2-s -\kappa\big) \leq t \leq  \frac{1}{2}\big(p_3^2+p_4^2-s +\kappa\big)
\end{equation}
where $\kappa$ is the K\"all\'en function
\begin{equation}\label{kaellen}
\kappa\left(s,p_3^2,p_4^2\right) \equiv
  \sqrt{s^2 + p_3^4 + p_4^4 - 2 (s\,p_3^2 + p_3^2\,p_4^2 + p_4^2\,s)}\,.
\end{equation}

Let us consider the partonic amplitude for the production of the two off-shell vector bosons $V_1 V_2$
\begin{equation}
\mathcal{S}(s,t,p_3^2,p_4^2) = S^{\mu \nu}(p_1,p_2,p_3)\,
\epsilon_3^{\mu}(p_3)^*\,\epsilon_4^{\nu}(p_4)^*\,,
\end{equation}
where $\epsilon_3$ and $\epsilon_4$ are the two polarisation vectors of $V_1$ and $V_2$ respectively.
In this notation,
we keep an overall factor $e^2$ implicit, where $e$ is the positron charge.

In order to calculate the partonic current $S^{\mu\nu}(p_1,p_2,p_3)$, we consider its
tensorial structure.
Lorentz invariance restricts it to be a linear combination of $17$ independent structures
\begin{align}
S^{\mu\nu}(p_1,p_2,p_3) &= 
          \bar{u}(p_2)\, \p_3 u(p_1) \left[\, F_1 \, p_1^\mu p_1^\nu + F_2\, p_1^\mu p_2^\nu 
          + F_3\, p_1^\mu p_3^\nu \, \right] \nonumber \\
    &+ \bar{u}(p_2)\, \p_3 u(p_1) \left[\, F_4 \, p_2^\mu p_1^\nu + F_5\, p_2^\mu p_2^\nu 
    + F_6\, p_2^\mu p_3^\nu  \, \right] \nonumber \\
    &+ \bar{u}(p_2)\, \p_3 u(p_1) \left[\, F_7 \, p_3^\mu p_1^\nu + F_8\, p_3^\mu p_2^\nu 
    + F_9\, p_3^\mu p_3^\nu  \, \right] \nonumber \\
    &+ \bar{u}(p_2)\, \gamma^\mu u(p_1) \left[\, F_{10}\, p_1^\nu + F_{11}\, p_2^\nu 
    + F_{12}\, p_3^\nu  \, \right] \nonumber \\
    &+ \bar{u}(p_2)\, \gamma^\nu  u(p_1) \left[\, F_{13}\, p_1^\mu + F_{14}\, p_2^\mu 
    + F_{15}\, p_3^\mu  \, \right]  \nonumber \\
    &+ \bar{u}(p_2)\, \gamma^\mu \p_3 \gamma^\nu u(p_1)\, F_{16} \nonumber \\
    &+ \bar{u}(p_2)\, \gamma^\nu \p_3 \gamma^\mu u(p_1)\, F_{17}\,,
\end{align}
where the form factors $F_1\,,...\,,F_{17}$ are scalar functions of the Mandelstam variables
$s,t,p_3^2,p_4^2$ and of the number of space-time dimensions $d$.
To further constrain $S^{\mu \nu}$, we choose the Landau gauge for the
electroweak vector bosons with the transversality condition
\begin{equation}
\epsilon_3\cdot p_3 = \epsilon_4 \cdot p_4 = 0\,, \label{gauge}
\end{equation}
and the sum over polarisations
\begin{align}
&\sum_{pol} (\epsilon_3^\mu)^* \epsilon_3^\nu = - g^{\mu \nu} + \frac{p_3^\mu p_3^\nu}{p_3^2}\,,\nonumber \\
&\sum_{pol} (\epsilon_4^\mu)^* \epsilon_4^\nu = - g^{\mu \nu} + \frac{p_4^\mu p_4^\nu}{p_4^2}\,. \label{polsum}
\end{align}

Imposing condition~\eqref{gauge} we can reduce the number of independent tensor structures
to ten~\cite{Denner:1988tv,Diener:1997nx}, which can be chosen as
\begin{align}
&T_1^{\mu \nu} =  \bar{u}(p_2)\, \p_3 u(p_1)\, p_1^\mu p_1^\nu\,, \qquad
T_2^{\mu \nu} =  \bar{u}(p_2)\, \p_3 u(p_1)\, p_1^\mu p_2^\nu\,, \nonumber \\
&T_3^{\mu \nu} =  \bar{u}(p_2)\, \p_3 u(p_1)\, p_2^\mu p_1^\nu\,, \qquad
T_4^{\mu \nu} =  \bar{u}(p_2)\, \p_3 u(p_1)\, p_2^\mu p_2^\nu\,, \nonumber \\
&T_5^{\mu \nu} =  \bar{u}(p_2)\, \gamma^\mu u(p_1)\, p_1^\nu \,, \qquad \hspace{0.32cm}
T_6^{\mu \nu} =  \bar{u}(p_2)\, \gamma^\mu u(p_1)\, p_2^\nu\,,\nonumber \\
&T_7^{\mu \nu} =  \bar{u}(p_2)\, \gamma^\nu u(p_1)\, p_1^\mu \,, \qquad \hspace{0.32cm}
T_8^{\mu \nu} =  \bar{u}(p_2)\, \gamma^\nu u(p_1)\, p_2^\mu\,,\nonumber \\
&T_9^{\mu \nu}  = \bar{u}(p_2)\, \gamma^\mu \p_3 \gamma^\nu u(p_1)\,, \qquad 
T_{10}^{\mu \nu}  = \bar{u}(p_2)\, \gamma^\nu \p_3 \gamma^\mu u(p_1)\,. \label{tensors}
\end{align}
Without any loss of generality we can thus write the partonic current as
\begin{align}
S^{\mu\nu}(p_1,p_2,p_3) &= \sum_{j=1}^{10}\, A_{j}(s,t,p_3^2,p_4^2)\, T_j^{\mu \nu} \,, \label{tensorstr}
\end{align}
where we introduced the new physical form factors $A_1,...,A_{10}$, which are again
scalar functions of the Mandelstam variables $s,t,p_3^2,p_4^2$ and of the dimension $d$.

Note that in deriving~\eqref{tensorstr} no assumption has been made on the dimensionality $d$, 
such that this decomposition
is valid for any continuous values of $d$. Its structure has been constrained using solely Lorentz
and gauge invariance and
is therefore true at every order in perturbation theory. On the other hand, 
the scalar coefficients $A_{j}(s,t,p_3^2,p_4^2)$
contain the explicit dependence on the perturbative order at which they are computed.
These coefficients can be extracted from the amplitude by applying appropriate projecting operators
on the latter. The projectors themselves can be expanded in the same tensorial basis:
\begin{align}
P_j^{\mu \nu} = \sum_{i=1}^{10} B_{ji}\, ( T_i^{\mu \nu} )^\dagger\,, \qquad j=1,10\,, \label{proj}
\end{align}
where the coefficients $B_{ji}$ are functions of the Mandelstam variables $s,t,p_3^2,p_4^2$ and
of the dimension $d$. They can be determined imposing that
\begin{equation}
\sum_{pol}\,P_j^{\mu_1 \mu_2}\,\left[  \epsilon_{3\mu_1}\,\epsilon_{4\,\mu_2}\,  
\epsilon_{3\nu_1}^*\,\epsilon_{4\,\nu_2}^*\, \right] S^{\nu_1 \nu_2} = A_j\,.
\end{equation}
Note that the contraction is performed in $d$ dimensions and at every stage 
one should always recall to use the polarisation sum in~\eqref{polsum}.
For later convenience we introduce also the following scalar quantities:
\begin{align}
\tau_i = \sum_{pol} \, \left(T_{i}^{\mu_1 \mu_2}\right)^\dagger \,\left[  \epsilon_{3\mu_1}\,\epsilon_{4\,\mu_2}\,  
\epsilon_{3\nu_1}^*\,\epsilon_{4\,\nu_2}^*\, \right] S^{\nu_1 \nu_2}\,, \label{tauj}
\end{align}
which are related to the coefficients $A_j$ according to
\begin{equation}
A_j = \sum_{i=1}^{10} B_{ji}\, \tau_i\,, \label{ajintauj}
\end{equation}
with the same coefficients $B_{ji}$ as in~\eqref{proj}.
These quantities (rather than the coefficients $A_j$) are particularly useful in order to build up the contractions of the 
$n$-loop amplitudes with the tree-level ones (see Appendix~\ref{sec:equalmass}).
We provide explicit expressions for $B_{ji}$ in computer readable format on \hepforge.

The partonic current receives contributions from QCD radiative corrections
and can be decomposed perturbatively as
\begin{equation}
S_{\mu \nu}(p_1,p_2,p_3) = S_{\mu \nu}^{(0)}(p_1,p_2,p_3) 
+ \left( \frac{\alpha_s}{2 \pi} \right)S_{\mu \nu}^{(1)}(p_1,p_2,p_3) 
+ \left( \frac{\alpha_s}{2 \pi} \right)^2S_{\mu \nu}^{(2)}(p_1,p_2,p_3)  
+ \mathcal{O}(\alpha_s^3)\,.
\end{equation}
Obviously also the un-renormalised tensor coefficients $A_j$ 
(or, equivalently the $\tau_j$) 
have the same perturbative expansion of the partonic amplitude
\begin{align}\label{Ajperturb}
A_j &= A_j^{(0)}
+ \left( \frac{\alpha_s}{2 \pi} \right)A_j^{(1)}
+ \left( \frac{\alpha_s}{2 \pi} \right)^2 A_j^{(2)} + \mathcal{O}(\alpha_s^3)\,,
\\
\tau_j &= \tau_j^{(0)}
+ \left( \frac{\alpha_s}{2 \pi} \right) \tau_j^{(1)}
+ \left( \frac{\alpha_s}{2 \pi} \right)^2 \tau_j^{(2)} + \mathcal{O}(\alpha_s^3)\,,
\end{align}
where the dependence on the Mandelstam variables is again implicit.

\section{\texorpdfstring
{Helicity amplitudes for $q\bar{q}' \to V_1 V_2\to \mathrm{4~leptons}$}
{Helicity amplitudes for qq' -> V1 V2 -> 4 leptons}
}
\label{sec:hel}
In physical applications we are interested in the processes
\begin{equation}
q(p_1) + \bar{q}'(p_2) \rightarrow V_1(p_3) + V_2(p_4) 
\rightarrow l_5(p_5) + \bar{l}_6(p_6) + l_7(p_7) + \bar{l}_8(p_8)
\end{equation}
where each of the two off-shell electroweak vector bosons can decay to pairs of
leptons, such that
$p_3= p_5+p_6$ and $p_4=p_7+p_8$.
Let us first focus on the general structure of the helicity amplitudes for this process.
Schematically these amplitudes can be written 
as the product of the partonic current 
 $S^{\mu \nu}$, and two leptonic currents $L_{\mu}\,,L_{\nu}$, mediated by the propagators 
 of the two off-shell vector bosons $P^{V}_{\mu \nu}(q)$
\begin{equation}
\widetilde{\M} (p_5,p_6,p_7,p_8; p_1,p_2) = S^{\mu \nu}(p_1,p_2,p_3)\,
P^{V_1}_{\mu \rho}(p_3)
L_{\rho}(p_5,p_6)\, P^{V_2}_{\nu \sigma}(p_4)\, L_\sigma(p_7,p_8)\,,
\end{equation}
where we stripped off electroweak couplings not relevant here
and postpone their discussion to the presentation of the full amplitude
in~\eqref{HelAmpl} below.
In the $R_\xi$-gauges the propagator of a vector boson $V$ reads
\begin{equation}
P_{\mu \nu}^V(q) = \frac{i\, \Delta^V_{\mu \nu}(q,\xi)}{D_V(q)}\,,
\end{equation}
with 
\begin{equation}
\Delta^V_{\mu \nu}(q,\xi) = \left( - g_{\mu \nu} + (1 - \xi) \frac{q_\mu q_\nu}{q^2 - \xi m_V^2}\right)\,,
\end{equation}
\begin{align}
&D_{\gamma^*}(q) = q^2\,,\qquad D_{Z,W}(q) = (q^2 - m_V^2 + i\,\Gamma_V m_V)\,,
\end{align}
where $m_V$ is the mass of the gauge boson and $\Gamma_V$ is its decay width.
While the Landau gauge used in the previous Section corresponds to $\xi\to 0$,
the term proportional to $(1-\xi)$ effectively vanishes for any $\xi$
since the electroweak vector bosons are directly coupled to massless fermion lines.

By fixing the helicities of the incoming partons and of the outgoing leptons
one sees that the left- and right-handed partonic production currents can be written as
\begin{align}
&S^{\mu \nu}_L(p_1^-,p_2^+,p_3) = \bar{v}_+(p_2) \Gamma^{\mu \nu} u_-(p_1) = 
 \langle 2\, | \Gamma^{\mu \nu} | \,1\, ]\,, \\
&S^{\mu \nu}_R(p_1^+,p_2^-,p_3) = \bar{v}_-(p_2) \Gamma^{\mu \nu} u_+(p_1) =
 [ 2\, | \Gamma^{\mu \nu} | \,1\, \rangle \,,
\end{align}
where the $\Gamma^{\mu \nu}$ are rank two-tensors and 
contain an odd number of $\gamma$-matrices.
We note in passing that, by complex conjugation, one gets
\begin{align*}
\left[ S^{\mu \nu}_R(p_1^+,p_2^-,p_3) \right]^* = \left( \, [ 2\, | \Gamma^{\mu \nu} | \,1\, \rangle\,  \right)^*
= \langle 2\, | \Gamma^{\mu \nu} | \,1\, ] = S^{\mu \nu}_L(p_1^-,p_2^+,p_3)  \,, 
\qquad \mbox{for all}
 \; \Gamma^{\mu \nu}\,.
\end{align*}
The left- and right-handed leptonic decay currents, on the other hand, can be written as
\begin{align}
&L^\mu_L(p_5^-,p_6^+) = \bar{u}_-(p_5) \,\gamma^{\mu} \, v_+(p_6) = [ 6\, | \gamma^{\mu } | \,5\, \rangle 
= \langle 5\, | \gamma^{\mu } | \,6\, ]  \,,\label{LepCurrL}\\
&L^\mu_R(p_5^+,p_6^-) = \bar{u}_+(p_5) \,\gamma^{\mu}\, v_-(p_6)=  [ 5\, | \gamma^{\mu } | \,6\, \rangle 
= \left( L^\mu_L(p_5^-,p_6^+) \right)^* = L_L^\mu(p_6^-,p_5^+)\,. \label{LepCurrR}
\end{align}

Note in particular that, as far as the lepton currents are concerned, a permutation of the external
momenta corresponds to a flip of the helicity.
All possible helicity amplitudes can be therefore obtained from the two basic amplitudes
\begin{align}
\M_{LLL}(p_1,p_2;p_5,p_6,p_7,p_8) &= S_L^{\mu \nu}(p_1^-,p_2^+,p_3)\,L_{L\mu}(p_5^-,p_6^+) L_{L\nu}(p_7^-,p_8^+)\,,\\
\M_{RLL}(p_1,p_2;p_5,p_6,p_7,p_8) &= S_R^{\mu \nu}(p_1^+,p_2^-,p_3)\,L_{L\mu}(p_5^-,p_6^+) L_{L\nu}(p_7^-,p_8^+)\,,
\end{align}
by simple permutations of the leptonic momenta. In particular we find
\begin{align}
&\M_{LLR}(p_1,p_2;p_5,p_6,p_7,p_8) = \M_{LLL}(p_1,p_2;p_5,p_6,p_8,p_7) \,, \nonumber \\
&\M_{LRL}(p_1,p_2;p_5,p_6,p_7,p_8) = \M_{LLL}(p_1,p_2;p_6,p_5,p_7,p_8) \,, \nonumber \\
&\M_{LRR}(p_1,p_2;p_5,p_6,p_7,p_8) = \M_{LLL}(p_1,p_2;p_6,p_5,p_8,p_7)\,, \nonumber \\
&\M_{RLR}(p_1,p_2;p_5,p_6,p_7,p_8) = \M_{RLL}(p_1,p_2;p_5,p_6,p_8,p_7) \,, \nonumber \\
&\M_{RRL}(p_1,p_2;p_5,p_6,p_7,p_8) = \M_{RLL}(p_1,p_2;p_6,p_5,p_7,p_8) \,, \nonumber \\
&\M_{RRR}(p_1,p_2;p_5,p_6,p_7,p_8) = \M_{RLL}(p_1,p_2;p_6,p_5,p_8,p_7)\,.
\end{align}

In order to put together the helicity amplitudes in their final form we need also to 
take into account the electroweak couplings of the gauge bosons to the partonic- and leptonic-currents
which we have been kept implicit so far.
We follow~\cite{Denner:1991kt} and parametrise the coupling of a 
vector boson $V$ to a fermion pair $f_1 f_2$ as

\begin{equation}
\mathcal{V}_\mu^{V f_1 f_2} = i\, e\, \Gamma_\mu^{V f_1 f_2}\,, \qquad 
\mbox{where} \quad e = \sqrt{4\,\pi\,\alpha} \quad \mbox{is the positron charge}\,, 
\end{equation}
such that all fermion charges are expressed in units of $e$ and 
\begin{equation}
\Gamma_\mu^{V f_1 f_2} = L_{f_1 f_2}^V \, \gamma_\mu \left( \frac{1-\gamma_5}{2}\right)
+ R_{f_1 f_2}^V \, \gamma_\mu \left( \frac{1+\gamma_5}{2}\right)\,.
\end{equation}
The left- and right-handed interactions are equal for a purely vectorial interaction.
Depending on the different kinds of gauge bosons we have
\begin{alignat}{2}
L_{f_1f_2}^\gamma &= -e_{f_1} \,\delta_{f_1f_2} &
R_{f_1f_2}^\gamma &= -e_{f_1} \,\delta_{f_1f_2}\,,
\label{gLBcoupl}
\\
L_{f_1f_2}^Z &= \frac{I_3^{f_1} - \sin^2 {\theta_w} e_{f_1}}{\sin{\theta_w} \cos{\theta_w}} \,\delta_{f_1f_2}\,, &\qquad
R_{f_1f_2}^Z &= -\frac{\sin{\theta_w} e_{f_1}}{\cos{\theta_w}} \,\delta_{f_1f_2}\,,
\label{ZLRcoupl}
\\
L_{f_1f_2}^W &= \frac{1 }{\sqrt{2}\, \sin{\theta_w}} \,\epsilon_{f_1f_2} \,,&
R_{f_1f_2}^W &= 0\,, \label{WLRcoupl}
\end{alignat}
where again
the charges $e_i$ are measured in terms of the fundamental electric charge $e>0$ and 
$\epsilon_{f_1f_2}$ is unity for $f_1\neq f_2$, but belonging to the same isospin doublet and respecting charge conservation, and zero otherwise.

Putting everything together
we find for the two independent helicity amplitudes for
$q\bar{q}' \to V_1 V_2 \to l_5 \bar{l}_6 l_7 \bar{l}_8$
\begin{align}
\mathcal{M}_{\lambda LL}^{V_1 V_2}(p_1, p_2;p_5,p_6,p_7,p_8) &= 
 (4 \pi \alpha)^2 \;
\frac{ L_{l_5 l_6}^{V_1} L_{l_7 l_8}^{V_2} }
{D_{V_1}(p_3)D_{V_2}(p_4)}\; \M_{\lambda LL}(p_1, p_2;p_5,p_6,p_7,p_8)\,,
\label{HelAmpl}
\end{align}
where $\lambda = L,R$ and we have bracketed out the tree-level dependence 
on the electric charge $i (4 \pi \alpha)^2$ and on the couplings with the decay lepton currents.
Obviously the corresponding helicity amplitudes for right-handed leptonic currents can be
obtained by the simple exchange $L_{f_i f_j}^{V} \leftrightarrow R_{f_i f_j}^V$ together with 
$p_i \leftrightarrow p_j$.

Once the tensor structure~\eqref{tensors} is given, we can perform
the contraction with the leptonic decay currents and 
fix the helicities of the incoming and outgoing fermions.
This enables us to cast the two independent helicity amplitudes $\M_{LLL}$
and $\M_{RLL}$ in the familiar spinor-helicity notation~\cite{Dixon:1996wi,Dixon:1998py}.
In doing so, one assumes that the external states are 4-dimensional 
and this allows to obtain one further Schouten identity between the 10 tensors structures,
such that one ends up with 9 independent form factors.
Our derivation is spelled out in detail in Appendix~\ref{sec:schouten}.
As a result, we obtain
\begin{align}
\M_{LLL}(p_1,p_2;p_5,p_6,p_7,p_8) &= 
 [ 1\, \p_3\, 2 \rangle\, \Big\{ 
   E_1\, \langle 15 \rangle \langle 17 \rangle [16][18] \nonumber \\
& 
+ E_2\, \langle 15 \rangle \langle 27 \rangle [16][28] 
+ E_3\, \langle 25 \rangle \langle 17 \rangle [26][18] \nonumber \\
& +
 E_4\, \langle 25 \rangle \langle 27 \rangle [26][28] \,
 + E_5 \langle 5 7 \rangle [ 68 ] \Big\}\nonumber \\
&+ E_6\, \langle 15 \rangle \langle 27 \rangle [16][18] 
+   E_7\, \langle 25 \rangle \langle 27 \rangle [26][18] \nonumber \\
&+ E_8\, \langle 25 \rangle \langle 17 \rangle [16][18] 
+ E_9\, \langle 25 \rangle \langle 27 \rangle [16][28]\,,   \label{MLLL}
\end{align}
\begin{align}
\M_{RLL}(p_1,p_2;p_5,p_6,p_7,p_8) &= 
 [ 2\, \p_3\, 1 \rangle\, \Big\{ 
   E_1\, \langle 15 \rangle \langle 17 \rangle [16][18] \nonumber \\
&
+ E_2\, \langle 15 \rangle \langle 27 \rangle [16][28] 
+ E_3\, \langle 25 \rangle \langle 17 \rangle [26][18] \nonumber \\
& +
 E_4\, \langle 25 \rangle \langle 27 \rangle [26][28] \,
 + E_5 \langle 5 7 \rangle [ 68 ] \Big\}\nonumber \\
&+ E_6\, \langle 15 \rangle \langle 17 \rangle [16][28] 
+   E_7\, \langle 25 \rangle \langle 17 \rangle [26][28] \nonumber \\
&+ E_8\, \langle 15 \rangle \langle 17 \rangle [26][18] 
+ E_9\, \langle 15 \rangle \langle 27 \rangle [26][28]\,,  \label{MRLL}
\end{align}
where
$$  [ 1\, \p_3\, 2 \rangle = [15] \langle 52 \rangle  + [16] \langle 62 \rangle \,, \qquad 
      [ 2\, \p_3\, 1 \rangle = [25] \langle 51 \rangle  + [26] \langle 61 \rangle\,,  $$
and the 9 form factors $E_j$ are linear combinations of the form factors $A_j$
\begin{align}
E_1 &= A_1\,,
&
E_6 &= 2 \, A_7 + \frac{2\,(u - p_3^2)}{s}\left(  A_9 - A_{10} \right)\,,
\nonumber \\
E_2 &= A_2 + \frac{2}{s}\left( A_9 - A_{10} \right)\,,
&
E_7 &= 2 \, A_8 - \frac{2\,(t - p_3^2)}{s}\left(  A_9 - A_{10} \right)\,,
\nonumber \\
E_3 &= A_3 - \frac{2}{s} \left( A_9 - A_{10} \right)\,,
&
E_8 &= 2 \, A_5 - \frac{2}{s} \left[ 
( u-s -p_3^2  ) A_9 + (t - p_4^2) A_{10} \right]\,,
\nonumber \\
E_4 &= A_4\,,
&
E_9 &=2 \, A_6 - \frac{2}{s} \left[ 
(t- s - p_3^2) A_{10} + (u - p_4^2) A_{9} \right]\,.
\nonumber \\
E_5 &= 2 \left( A_9 + A_{10} \right)\,.
&&
\label{EjinAj}
\end{align}
In the following, we will consider a perturbative expansion of the form factors
$E_j$ defined completely analogous to that of the coefficients $A_j$ in \eqref{Ajperturb}.
We note that the expressions~\eqref{MLLL} and~\eqref{MRLL} are formally identical to the
corresponding formulas derived in~\cite{Caola:2014iua}, such that our form factors $E_j$
can be mapped one to one to the $F_j$ defined in~\cite{Caola:2014iua}.

In the next Section we will describe how to compute form factors $A_1,\ldots,A_{10}$ and therefore
also form factors $E_1,\ldots,E_9$
at tree-level, one-loop and two-loop order, following a straightforward diagrammatic
approach. 
In particular, we will discuss the different electroweak coupling arrangements $\C$
contributing to the functions $A_j$ and $E_j$,%
\begin{align}\label{AdecompEW}
A_j &= \delta_{i_1 i_2} \sum_{\C} Q^{\lambda,V_1V_2,[\C]}_{q\,q'} A_j^{[\C]},
\qquad j = 1,\ldots,10,\notag\\
E_j &= \delta_{i_1 i_2} \sum_{\C} Q^{\lambda,V_1V_2,[\C]}_{q\,q'} E_j^{[\C]},
\qquad j = 1,\ldots,9,
\end{align}
\\[-1.15em]
where $Q^{\lambda,V_1V_2,[j]}_{q\,q'}$ denotes a coupling factor,
$\lambda$ is the helicity of the incoming quark,
and $i_1$, $i_2$ are the colours of the incoming quark and anti-quark,
respectively.

We want to stress once more an important point. Reducing the 
10 coefficients $A_j$ to the 9 coefficients $E_j$ required the assumption that the external 
states can be treated as 4-dimensional. 
In order to avoid any loss of information, we will work considering the $A_j$ as fundamental objects 
(derived in $d$ dimensions throughout)
and refer to formulas~\eqref{EjinAj} in order to reconstruct the $E_j$ explicitly.

\section{Organisation of the calculation}
\label{sec:calc}

The calculation of the two-loop helicity
amplitudes can be set up in a way that is independent on the nature of the vector bosons considered, 
by organising the Feynman diagrams contributing to any such process into different classes.
We find in particular that, as long as we limit ourselves to QCD corrections only,
at any given number of loops, seven different types of diagrams 
can contribute, depending on the arrangement of the external vector bosons.

\begin{figure}
\centerline{
\includegraphics[width=0.85\textwidth]{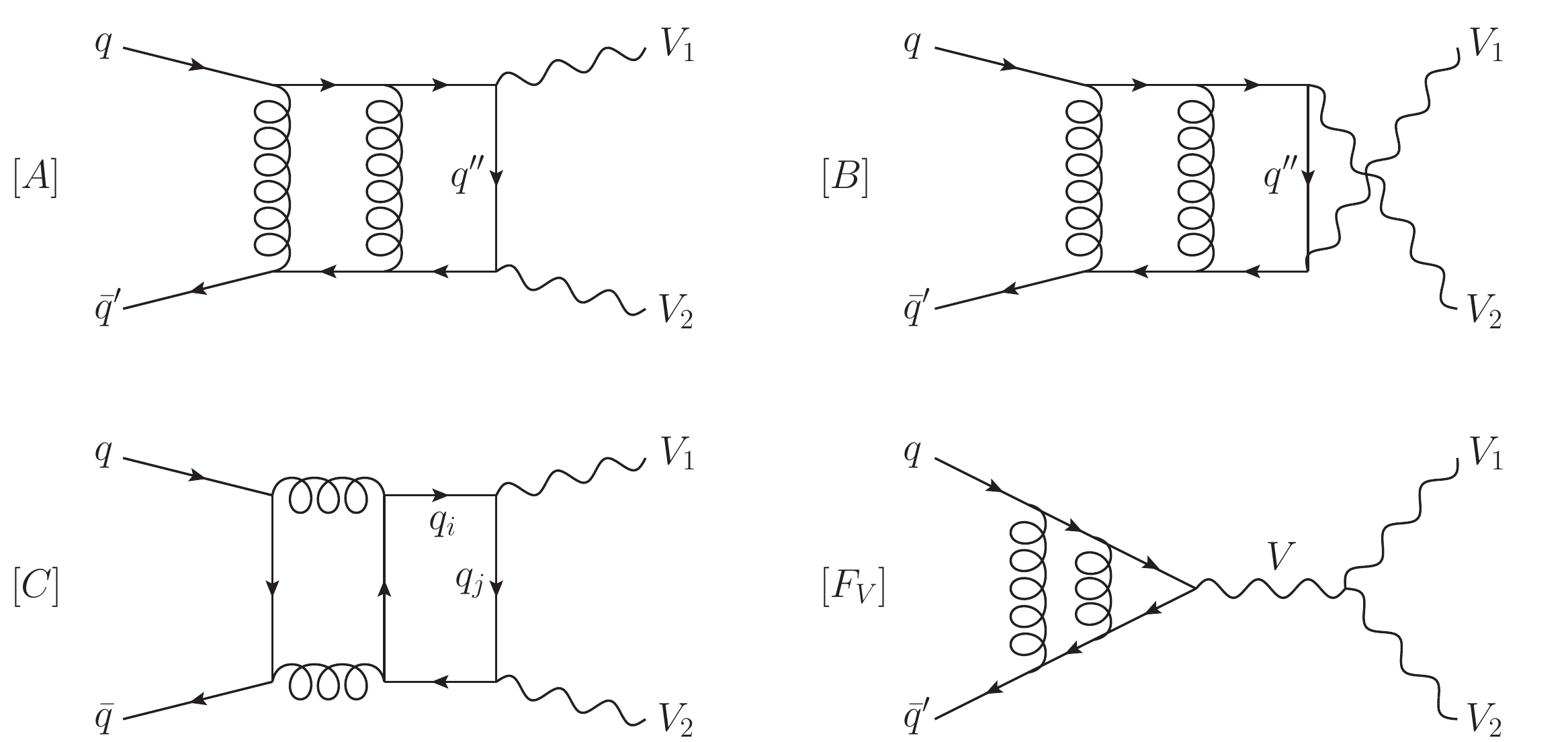}
}
\caption{\label{fig:diagclass}
Representative Feynman diagrams for classes $A$, $B$, $C$ and $F_V$
relevant for the production of two electroweak vector bosons at the two-loop level.
All of these classes receive contributions both from planar and non-planar diagrams.
}
\end{figure}

\begin{description}
\item[Class $\mathbf{A}$] collects all those diagrams where both vector bosons are attached on the external fermion line, 
such that $V_1$ is adjacent to the quark $q(p_1)$. In the case of a left-handed (right-handed) quark amplitude
these diagrams are proportional to $L_{q\,q''}^{V_1}\,L_{q''q'}^{V_2}$ 
($R_{q\,q''}^{V_1}\,R_{q''q'}^{V_2}$).

\item[Class $\mathbf{B}$] collects all those diagrams where both vector bosons are attached on the external fermion line, 
such that $V_1$ is adjacent to 
the antiquark $\bar{q}'(p_2)$. Also these diagrams, in the case of a left-handed (right-handed) quark amplitude,
are proportional to $L_{q'q''}^{V_1}\,L_{q''q}^{V_2}$ 
($R_{q'q''}^{V_1}\,R_{q''q}^{V_2}$).

\item[Class $\mathbf{C}$] contains instead all diagrams where both vector bosons are attached to a fermion loop.
These diagrams are proportional to the charge weighted sum of the quark flavours, which we denote as
$N_{V_1 V_2}$, depending on nature of the final state bosons. 
In the general case, these diagrams yield two different contributions.
In the first one, which is proportional to the sum of the vector-vector and the axial-axial couplings,
all dependence on $\gamma_5$ cancels out. The vector-axial contribution, instead, is linear in $\gamma_5$. 
Nevertheless, this last contribution is expected to always vanish identically for massless 
quarks running in the loops, for any choice of $V_1$ and $V_2$, 
due to charge parity conservation~\cite{Glover:1988rg,Glover:1988fe,Caola:2014iua,Melnikov:2015laa}.
Taking this into account we find
\begin{align}\label{coupl1}
N_{\gamma \gamma} &= \frac{1}{2}\sum_{i} \left[ \left( L_{q_i q_i}^\gamma \right)^2 + \left( R_{q_i q_i}^\gamma \right)^2 \right], &
N_{Z \gamma} &=  \frac{1}{2} \sum_{i} \left( L_{q_i q_i}^Z L_{q_i q_i}^\gamma + R_{q_i q_i}^Z R_{q_i q_i}^\gamma \right), \nonumber\\
N_{ZZ} &= \frac{1}{2} \sum_{i} \left[  \left( L_{q_i q_i}^Z\right)^2 + \left(R_{q_i q_i}^Z \right)^2 \right], &
N_{WW} &= \frac{1}{2} \sum_{i,\,j} \left( L_{q_i q_j}^W  L_{q_j q_i}^W \right),
\end{align}
where the indices $i,j$ run over the flavours of the quarks in the loop
and $L_{q_i q_i}^\gamma = R_{q_i q_i}^\gamma$.
Of course, $N_{\gamma\gamma}=\sum_{i} e_{q_i}^2$ and, due to charge conservation,
$N_{W \gamma} = N_{W Z} = 0\,.$

\item[Class $\mathbf{D_1}$] contains all diagrams where $V_1$ is attached to a fermion loop and $V_2$ to the external
fermion line. Up to two loops, the diagrams in this class must sum up to zero due to Furry's theorem.
\item[Class $\mathbf{D_2}$] contains all diagrams where $V_2$ is attached to a fermion loop and $V_1$ to the external
fermion line. At two loops the diagrams in this class, as for the previous case, must sum up to zero due to Furry's theorem.
\item[Class $\mathbf{E}$] contains all diagrams there $V_1$ and $V_2$ are attached to two different fermion loops.
These diagrams contribute only starting from three-loop order and we can ignore them.
\item[Classes $\mathbf{F_V}$] collect the form-factor diagrams where the production of 
the two vector bosons $V_1,V_2$ is mediated by the exchange of another vector boson $V$.
Depending on the type of vector bosons $V_1,V_2$ there can be more than one such class
due to different intermediate vector bosons.
In the case of a left-handed (right-handed) quark amplitude these diagrams are proportional to
$L_{q\,q'}^V\, c_{V\, V_1 V_2}$ ($R_{q\,q'}^V\, c_{V\, V_1 V_2}$), where $c_{V\, V_1 V_2}$
is the electroweak coupling of the triple gauge boson vertex defined
for all particles and momenta outgoing as
\begin{align}
\mathcal{V}_{V\, V_1 V_2}^{\rho \mu \nu}(a,b,c) = i\,e\,c_{V\,V_1V_2}\, \left[ \, (a-b)^\nu g^{\mu \rho}
+ (b-c)^\rho g^{\mu \nu} + (c-a)^\mu g^{\nu \rho}\,
\right]
\end{align}
with
\begin{align} \label{coupl2}
  c_{\gamma W^\pm W^\mp} &= c_{W^\mp\gamma W^\pm} = c_{W^\pm W^\mp\gamma} = \pm 1\,,
\notag\\ 
  c_{Z W^\pm W^\mp} &= c_{W^\mp Z W^\pm} = c_{W^\pm W^\mp Z} = \mp \cot\theta_w
\,.
\end{align}
\end{description}

It is clear that, depending on the nature of the vector bosons $V_1, \,V_2$ and on the loop order,
not all classes above will give non-zero contribution. 
At tree-level, for example, only classes $A$, $B$ and $F_V$
can contribute. The same is true also at one loop, provided that we limit ourselves to QCD corrections only.
The situation changes at two loops, where also diagrams for classes $C,\,D_1$ and $D_2$ occur.
Notice moreover that the form-factor type diagrams in
class $F_V$ are relevant only for the production of $W\gamma$, $W Z$ or $WW$ pairs.

Up to two loops, we can thus restrict the summation in  (\ref{AdecompEW}) to $\C=A,B,C,F_V$.
We show representative diagrams in Figure~\ref{fig:diagclass}.
For the coupling factors we have
\begin{align} 
Q^{L,V_1V_2,[A]}_{q\,q'}  & =  L_{q\,q''}^{V_1}\,L_{q''q'}^{V_2}\;, &
Q^{R,V_1V_2,[A]}_{q\,q'}  & =  R_{q\,q''}^{V_1}\,R_{q''q'}^{V_2}\;,  \nonumber \\
Q^{L,V_1V_2,[B]}_{q\,q'}  & =  L_{q'q''}^{V_1}\,L_{q''q}^{V_2}\;,&
Q^{R,V_1V_2,[B]}_{q\,q'}  &=  R_{q'q''}^{V_1}\,R_{q''q}^{V_2}\;,  \nonumber \\
Q^{L,V_1V_2,[C]}_{q\,q'}  & = N_{V_1V_2} \delta_{q\,q'}\;, &
Q^{R,V_1V_2,[C]}_{q\,q'}  &= N_{V_1V_2} \delta_{q\,q'} \;, \nonumber \\
Q^{L,V_1V_2,[F_V]}_{q\,q'}  & =  \frac{L_{q\,q'}^V c_{VV_1V_2}}{s-m_V^2-i\,\Gamma_V\,m_V} \;, &
Q^{R,V_1V_2,[F_V]}_{q\,q'}  & =  \frac{R_{q\,q'}^V c_{VV_1V_2}}{s-m_V^2-i\,\Gamma_V\,m_V} \;. \label{coupl3}
\end{align}
With these definitions, the value of the coefficients $A_j^{[F_V],(n)}$ do not
depend on the nature of the mediating vector boson $V$, such that in particular
\begin{equation}
A_j^{[F_\gamma],(n)} = A_j^{[F_Z],(n)}= A_j^{[F_W],(n)} = A_j^{[F],(n)}\,.
\end{equation}

We have computed the coefficients $A_j$ for the different classes of diagrams contributing
at tree level, one loop and two loops, namely $A_j^{[\C],(0)}$,  $A_j^{[\C],(1)}$,  $A_j^{[\C],(2)}$,
with $\C = A,B,C,D_1,D_2,F$.

At tree-level order we find
\begin{align}
A_7^{[A],(0)} &= -\frac{2}{t}\,,&
  A_{10}^{[A],(0)} &= +\frac{1}{t}\,, &
  A_j^{[A],(0)} &= 0\,, \quad j=1,...,6,8,9\,, \nonumber \\
A_8^{[B],(0)} &= +\frac{2}{u}\,, &
  A_9^{[B],(0)} &= -\frac{1}{u}\,, &
  A_j^{[B],(0)} &= 0\,, \quad j=1,...,7,10\,, \nonumber \\
A_7^{[F],(0)} &= A_8^{[F],(0)} = +2\,, &
  A_9^{[F],(0)} &= A_{10}^{[F],(0)} = -1\,, &
  A_j^{[F],(0)} &= 0\,, \quad j=1,...,6\,. \label{HelAmplTree}
\end{align}

We can notice immediately that, as far as the form-factor type diagrams are concerned,
any $n$-loop QCD corrections will not modify the structure of~\eqref{HelAmplTree}, and in particular
we have
\begin{align}
A_j^{[F],(n)} &= \F^{(n)}(s) A_j^{[F],(0)} \label{HelAmplFV}
\end{align}
where $\F^{(n)}(s)$ are the $n$-loop QCD corrections to the quark
form-factor.

Let us discuss the features of our $E_j$ set of coefficients, which is relevant
for the four-dimensional helicity amplitudes for the full $2\to4$ process.
We consider crossings of external legs described by the permutations
\begin{align}
&\pi_{12} := p_1 \leftrightarrow p_2 \Rightarrow \{\,t \leftrightarrow u \,\} \nonumber \\
&\pi_{34} := p_3 \leftrightarrow p_4 \Rightarrow \{\, t \leftrightarrow u\,, \;\; p_3^2 \leftrightarrow p_4^2 \,\} \,.
\end{align}
and focus on the behaviour of the $E_j^{[\C]}$ for the non-trivial cases $\C = A,B,C$.
From the exchange of quark and anti-quark, $\pi_{12}$ we find for the
amplitudes
\begin{align}
\M^{[A]}_{LLL} =  - \M^{[B]}_{RLL}(p_1 \leftrightarrow p_2)\,, \quad \M^{[C]}_{LLL} =  - \M^{[C]}_{RLL}(p_1 \leftrightarrow p_2)\,,
\end{align}
from which one can directly obtain
\begin{align}
E_1^{[A]}(s,t,p_3^2,p_4^2) &= - E_4^{[B]}(s,u,p_3^2,p_4^2)\,,
&
E_8^{[A]}(s,t,p_3^2,p_4^2) &= - E_9^{[B]}(s,u,p_3^2,p_4^2)\,, 
\nonumber \\
E_2^{[A]}(s,t,p_3^2,p_4^2) &= - E_3^{[B]}(s,u,p_3^2,p_4^2)\,,
&
E_9^{[A]}(s,t,p_3^2,p_4^2) &= - E_8^{[B]}(s,u,p_3^2,p_4^2)\,,
\nonumber \\
E_3^{[A]}(s,t,p_3^2,p_4^2) &= - E_2^{[B]}(s,u,p_3^2,p_4^2)\,,
&
E_1^{[C]}(s,t,p_3^2,p_4^2) &= - E_4^{[C]}(s,u,p_3^2,p_4^2)\,,
\nonumber \\
E_4^{[A]}(s,t,p_3^2,p_4^2) &= - E_1^{[B]}(s,u,p_3^2,p_4^2)\,,
&
E_2^{[C]}(s,t,p_3^2,p_4^2) &= - E_3^{[C]}(s,u,p_3^2,p_4^2)\,,
\nonumber \\
E_5^{[A]}(s,t,p_3^2,p_4^2) &= - E_5^{[B]}(s,u,p_3^2,p_4^2)\,,
&
E_5^{[C]}(s,t,p_3^2,p_4^2) &= - E_5^{[C]}(s,u,p_3^2,p_4^2)\,,
\nonumber \\
E_6^{[A]}(s,t,p_3^2,p_4^2) &= - E_7^{[B]}(s,u,p_3^2,p_4^2)\,,
&
E_6^{[C]}(s,t,p_3^2,p_4^2) &= - E_7^{[C]}(s,u,p_3^2,p_4^2)\,,
\nonumber \\
E_7^{[A]}(s,t,p_3^2,p_4^2) &= - E_6^{[B]}(s,u,p_3^2,p_4^2)\,, 
&
E_8^{[C]}(s,t,p_3^2,p_4^2) &= - E_9^{[C]}(s,u,p_3^2,p_4^2)\,, 
\label{Ejx12}
\end{align}
From exchange of the external vector bosons, $\pi_{34}$, we have
\begin{align}
\M^{[A]}_{\lambda LL} = \M^{[B]}_{\lambda LL}(p_3 \leftrightarrow p_4)\,, \quad \M^{[C]}_{\lambda LL} =  \M^{[C]}_{\lambda LL}(p_3 \leftrightarrow p_4)\,,
\quad \mbox{with} \quad \lambda = L,R\,,
\end{align}
which implies
\begin{align}
E_1^{[A]}(s,t,p_3^2,p_4^2) &= - E_1^{[B]}(s,u,p_4^2,p_3^2)\,,
&
E_9^{[A]}(s,t,p_3^2,p_4^2) &= + E_7^{[B]}(s,u,p_4^2,p_3^2)\,,
\nonumber\\
E_2^{[A]}(s,t,p_3^2,p_4^2) &= - E_3^{[B]}(s,u,p_4^2,p_3^2)\,,
&
E_1^{[C]}(s,t,p_3^2,p_4^2) &= - E_1^{[C]}(s,u,p_4^2,p_3^2)\,,
\nonumber\\
E_3^{[A]}(s,t,p_3^2,p_4^2) &= - E_2^{[B]}(s,u,p_4^2,p_3^2)\,,
&
E_2^{[C]}(s,t,p_3^2,p_4^2) &= - E_3^{[C]}(s,u,p_4^2,p_3^2)\,,
\nonumber\\
E_4^{[A]}(s,t,p_3^2,p_4^2) &= - E_4^{[B]}(s,u,p_4^2,p_3^2)\,,
&
E_4^{[C]}(s,t,p_3^2,p_4^2) &= - E_4^{[C]}(s,u,p_4^2,p_3^2)\,,
\nonumber\\
E_5^{[A]}(s,t,p_3^2,p_4^2) &= - E_5^{[B]}(s,u,p_4^2,p_3^2)\,,
&
E_5^{[C]}(s,t,p_3^2,p_4^2) &= - E_5^{[C]}(s,u,p_4^2,p_3^2)\,,
\nonumber\\
E_6^{[A]}(s,t,p_3^2,p_4^2) &= + E_8^{[B]}(s,u,p_4^2,p_3^2)\,,
&
E_6^{[C]}(s,t,p_3^2,p_4^2) &= + E_8^{[C]}(s,u,p_4^2,p_3^2)\,,
\nonumber\\
E_7^{[A]}(s,t,p_3^2,p_4^2) &= + E_9^{[B]}(s,u,p_4^2,p_3^2)\,,
&
E_7^{[C]}(s,t,p_3^2,p_4^2) &= + E_9^{[C]}(s,u,p_4^2,p_3^2)\,,
\nonumber\\
E_8^{[A]}(s,t,p_3^2,p_4^2) &= + E_6^{[B]}(s,u,p_4^2,p_3^2)\,,
&
&\phantom{= + E_6^{[B]}(s,u,p_4^2,p_3^2)}
\label{Ejx34}
\end{align}
Similar but slightly more involved relations can be derived for the primary
set of coefficients, $A_j$, but we don't list them here for brevity.
We have explicitly verified that relations~\eqref{Ejx12},\eqref{Ejx34} for the
$E_j$ and the corresponding relations for the $A_j$ hold for
our results at tree level, one loop and two loops.

While most of the coefficients $A_j$ are zero at tree level, fewer of the $E_j$
have this property.
We find for class A
\begin{align}
\label{HelAmplTreeEA}
E_1^{[A],(0)} &= 0,&
E_2^{[A],(0)} &= -\frac{2}{s t},&
E_3^{[A],(0)} &= \frac{2}{s t},\nonumber\\
E_4^{[A],(0)} &= 0,&
E_5^{[A],(0)} &= \frac{2}{t},&
E_6^{[A],(0)} &= -2 \frac{(s - t + p_4^2)}{s t},\nonumber\\
E_7^{[A],(0)} &= 2 \frac{(t - p_3^2)}{s t},&
E_8^{[A],(0)} &= -2 \frac{(t - p_4^2)}{s t},&
E_9^{[A],(0)} &= 2 \frac{(s - t + p_3^2)}{s t},
\\
\intertext{for class B}
\label{HelAmplTreeEB}
E_1^{[B],(0)} &= 0,&
E_2^{[B],(0)} &= -\frac{2}{s u},&
E_3^{[B],(0)} &= \frac{2}{s u},\nonumber\\
E_4^{[B],(0)} &= 0,&
E_5^{[B],(0)} &= -\frac{2}{u},&
E_6^{[B],(0)} &= -2 \frac{(u - p_3^2)}{s u},\nonumber\\
E_7^{[B],(0)} &= 2 \frac{(s - u + p_4^2)}{s u},&
E_8^{[B],(0)} &= -2 \frac{(s - u + p_3^2)}{s u},&
E_9^{[B],(0)} &= 2 \frac{(u - p_4^2)}{s u},
\\
\intertext{and for class F}
\label{HelAmplTreeEF}
E_1^{[F],(0)} &= 0,&
E_2^{[F],(0)} &= 0,&
E_3^{[F],(0)} &= 0,\nonumber\\
E_4^{[F],(0)} &= 0,&
E_5^{[F],(0)} &= -4,&
E_6^{[F],(0)} &= +4,\nonumber\\
E_7^{[F],(0)} &= +4,&
E_8^{[F],(0)} &= -4,&
E_9^{[F],(0)} &= -4.
\end{align}
As discussed above, class C contributions enter only at the two-loop level.

The calculation of the coefficients $A_j$ and thus $E_j$ proceeds as follows.
The diagrams belonging to class $F_V$ are known~\cite{Gehrmann:2005pd}. 
They do not have to be recomputed and we will
not refer to them anymore here.
As far as the other classes are concerned, 
we produced all the tree-level, one-loop and two-loop Feynman diagrams with \qgraf~\cite{Nogueira:1991ex}.
The scalar coefficients $A_j$ are evaluated analytically diagram by diagram by applying 
the projectors defined in~\eqref{proj} and summing over the polarisations of the external
vector bosons as in~\eqref{polsum}.
For the gluons we employ the Feynman-'t\,Hooft gauge.
All these manipulations are consistently performed in
$d$ dimensions.
Upon doing this we obtain the coefficients in terms of
a large number of scalar two-loop Feynman integrals. The latter are classified into three
integral families, two planar and one non-planar.
We have made 
use of \reduze\;2~\cite{vonManteuffel:2012np,Studerus:2009ye,Bauer:2000cp,fermat} in order to map
all integrals to these integral families and to perform a full integration-by-parts
reduction~\cite{Tkachov:1981wb,Chetyrkin:1981qh,Gehrmann:1999as,Laporta:2001dd}
of the latter to a set of master integrals.
All intermediate algebraic manipulations on the Feynman diagrams 
have been performed using \form~\cite{Vermaseren:2000nd}.
Once the coefficients $A_j$ for the different classes of diagrams are known at the different loop orders,
one can calculate the form factors $E_j$ using~\eqref{EjinAj}. 
Since the expressions for the coefficients $A_j$ (and equivalently those for the $E_j$) 
at two loops are very lengthy we decided not to include them explicitly in the text. 
Analytical expressions for the $A_j$, 
prior to UV renormalisation and IR subtraction, expressed as linear combinations of masters
integrals and retaining full dependence on the dimensions $d$ are available on
our project page at \hepforge.

\section{Master integrals}
\label{sec:masters}

\subsection{Computation via differential equations}
\label{sec:tradpolylogs}

We computed all two-loop master integrals needed for our process with the
method of differential
equations~\cite{Kotikov:1990kg,Remiddi:1997ny,Caffo:1998du,Gehrmann:1999as}
and optimised the solutions for fast and
precise numerical evaluations~\cite{Duhr:2012fh,Bonciani:2013ywa,Gehrmann:2014bfa}.
The master integrals for the case $p_3^2=p_4^2$ have first been calculated
in \cite{Gehrmann:2013cxs,Gehrmann:2014bfa}.
Here, we consider the case $p_3^2\neq p_4^2$, for which the master integrals
have been computed in \cite{Henn:2014lfa,Caola:2014lpa} for the first time.
Our calculation provides an independent check of these results
and improves them for numerical applications.
In this Section we present our calculation and discuss qualitative aspects
of the results.
We provide the explicit solutions in computer readable format on \hepforge.

We find that all master integrals are described by the integral families
presented in \cite{Gehrmann:2014bfa} for the case $p_3^2\neq p_4^2$
and crossings thereof.
We start by determining a set of linearly independent master integrals
for all relevant topologies using \reduze\;2~\cite{vonManteuffel:2012np}.
For convenience, we stick to the normal form definitions for the master
integrals given in \cite{Henn:2014lfa,Caola:2014lpa}.
We supplement these definitions by new normal form definitions for eight
factorisable topologies corresponding to products of one-loop integrals.
All our definitions are supplied in computer readable form on \hepforge.

We consider the master integrals of all integral families at the same time
and eliminate multiple variants of equivalent master integrals using the
shift-finder of \reduze\;2.
For this purpose we also identify crossed topologies and work out
relations between crossed and uncrossed master integrals.
Ignoring crossed variants and counting product topologies as two-loop
topologies we find a total number of 84 independent master integrals.
To apply the method of differential equations, we include also
crossed versions for a couple of integrals, which appear in sub-topologies
of non-planar topologies.
In this way we assemble a minimal set of 111 master integrals suitable
for the construction of a system of differential equations.

We compute the partial derivatives of the master integrals with respect to
all independent external invariants $s$, $t$, $p_3^2$, $p_4^2$ in terms of
master integrals with the help of \reduze.
The coefficients contain rational functions of the invariants and
the K\"all\'en function $\kappa$,~\eqref{kaellen}, associated to the two-body
phase space.

To rationalise the root $\kappa$, we employ the parametrisation
\begin{align}\label{n34params}
s &= \mm^2 (1+\xx)^2, &
p_3^2 &= \mm^2 \xx^2 (1-\yy^2),\nonumber\\
t &= -\mm^2 \xx ((1+\yy)(1+\xx \yy) - 2 \zz \yy (1+\xx)), &
p_4^2 &= \mm^2 (1 - \xx^2 \yy^2)\,,
\end{align}
(see eq.~(2.9) of \cite{Caola:2014lpa}).
In this parametrisation, we define the vector of master integrals
$\vec{M}=(M_i)$, $i=1,\ldots,111$, using the integral measure
\begin{equation}\label{measure}
 \left(\frac{C_\epsilon}{16 \pi^2}\right)^{-2}\, (\mm^2)^{2 \epsilon }
\int \frac{d^d k}{(2 \pi)^d}\frac{d^d l}{(2 \pi)^d}
\end{equation}
which absorbs the overall mass dimension $\mm$.
Here, $d = 4-2\epsilon$ and
\begin{equation}
  C_\epsilon =  (4 \pi)^\epsilon \, \frac{\Gamma(1+ \epsilon) \, 
\Gamma^2(1-\epsilon)}{\Gamma(1-2\epsilon)}\,.
\end{equation}
In the following, we will work directly in the physical region of phase space.
Due to the specific choice of the master integrals~\cite{Henn:2013pwa,Kotikov:2010gf},
the partial differential
equations combine into the simple total differential,
\begin{equation}
\ud \vec{M}(\epsilon; \xx, \yy, \zz) =
\epsilon \sum_{k=1}^{20} A_k \ud\ln(\bar{l}_k)\, \vec{M}(\epsilon; \xx, \yy, \zz)
\end{equation}
where the matrices $A_k$ contain just rational numbers and
the alphabet is
\begin{align} \label{n34alphabet}
\{ \bar{l}_1,\ldots,\bar{l}_{20} \} = \{ &
 2,
 \xx, 1 + \xx, 1 - \yy, \yy, 1 + \yy, 1 - \xx \yy, 1 + \xx \yy, 1 - \zz, \zz, 
 \notag\\ &
 1 + \yy - 2 \yy \zz, 1 - \yy + 2 \yy \zz,
 1 + \xx \yy - 2 \xx \yy \zz,
 1 - \xx \yy + 2 \xx \yy \zz, 
 \notag\\ &
 1 + \yy + \xx \yy + \xx \yy^2 - 2 \yy \zz - 2 \xx \yy \zz, 
 1 + \yy - \xx \yy - \xx \yy^2 - 2 \yy \zz + 2 \xx \yy \zz,
 \notag\\ &
 1 - \yy - \xx \yy + \xx \yy^2 + 2 \yy \zz + 2 \xx \yy \zz,
 1 - \yy + \xx \yy - \xx \yy^2 + 2 \yy \zz - 2 \xx \yy \zz, 
 \notag\\ &
 1 - 2 \yy - \xx \yy + \yy^2 + 2 \xx \yy^2 - \xx \yy^3 + 4 \yy \zz + 2 \xx \yy \zz + 2 \xx \yy^3 \zz,
 \notag\\ &
 1 - \yy - 2 \xx \yy + 2 \xx \yy^2 + \xx^2 \yy^2 - \xx^2 \yy^3 + 2 \yy \zz + 4 \xx \yy \zz +  2 \xx^2 \yy^3 \zz
\}\,.
\end{align}
Anticipating the solution, we included the letter 2 already, which is
of course arbitrary at the level of the differential equations.
While we found that it is possible to reduce the number of letters by
forming appropriate ratios, a reduction of the alphabet is best performed
using a different parametrisation, as we will see below.

After expansion in $\epsilon$ it is straight-forward to integrate the
differential equations in terms of multiple polylogarithms
\begin{equation}
\G(w_1,\dots,w_n; z) = \int_0^z \frac{\ud t}{t - w_1} \G(w_2,\dots,w_n;t),
\end{equation}
with $\G(0,\dots,0;z) = \frac{1}{n!}\ln^n(z)$ for $n$ zero weights and
$\G(;z)=1$.
For each order in $\epsilon$, we integrate the partial derivatives in $\zz$.
This gives the solution up to a function of $\xx$ and $\yy$.
We employ the partial derivatives in $\xx$ to determine this
function, this time up to a function of $\yy$.
Subsequent usage of the derivative in $\yy$ fixes the boundary terms
up to one constant per master integral for the given order in $\epsilon$.
Despite the presence of nonlinearities in~\eqref{n34alphabet}, the specific
order of our integrations ensures that in fact only \emph{linear} denominators
occur in the respective integration variable.
We integrate the master integrals through to weight 4, which corresponds
to $\epsilon^4$ terms in the chosen normalisation.
The necessary argument-change transformations for the multiple polylogarithms
were derived using an in-house package, which employs fitting of constants
using high precision samples obtained with~\cite{Vollinga:2004sn}.

In order to fix the integration constants, we consider the equal mass
limit $p_4^2 \to p_3^2$ which implies $\xx \to 1$.
This limit is smooth and our master integrals become simple linear
combinations of the normal form integrals defined in~\cite{Gehrmann:2014bfa},
where the coefficients in this map are just rational numbers.
We compute the limit at the level of our solutions and equate them
to the real-valued solutions of \cite{Gehrmann:2014bfa}.
Using the coproduct-augmented symbol calculus~\cite{2001math.3059G,Brown:2008um,Goncharov:2010jf,Duhr:2011zq,Duhr:2012fh,Duhr:2014woa},
we find perfect agreement for all non-constant terms and
easily fix the boundary constants of the present integrals.
We also compared our results to the solutions of
\cite{Henn:2014lfa,Caola:2014lpa} and find perfect agreement at the
analytical level.

The solutions we obtained in this way are not ideal for our purposes yet,
since their numerical evaluation is rather slow.
Moreover, they contain spurious structures: the individual
multiple polylogarithms contribute letters
$\{1 - \xx, 1 + \xx \yy^2, 2 + \xx + \xx \yy^2, 1 + 2 \xx + \xx^2 \yy^2\}$
which cancel for the master integral itself.
In particular, the equal-virtuality limit $\xx\to 1$ is completely smooth as can
be seen from \eqref{n34alphabet}, but the representation does not allow
for an evaluation exactly in the equal-virtuality point.

\subsection{Optimisation of the functional basis}
\label{sec:optpolylogs}

We wish to cast our solutions to a new representation which allows for
fast and stable numerical evaluations and is free of spurious letters.
In order to achieve that goal we select a new basis of multiple polylogarithms
where we do not force individual variables into the argument of $\G$ functions
anymore.
As a side effect, this gives us more freedom for a rational parametrisation,
since we avoid problems due to non-linear denominators in the integration variable.
It is convenient to choose new variables $x$, $y$, $z$ and $m^2$ according to
\begin{equation}\label{n12params}
 s =  m^2 (1+x) (1+x y),\quad
 t = -m^2 x z,\quad
 p_3^2 = m^2,\quad
 p_4^2 = m^2 x^2 y
\end{equation}
(see eq.~(2.7) of \cite{Caola:2014lpa}), which again rationalises the root $\kappa$.
We select the branch for which in the physical domain
\begin{equation}
x > 0,\quad  0 < y < z < 1,\quad m^2 > 0.
\end{equation}
This reparametrisation is actually not crucial for what will follow, but it decreases
the number of irreducible polynomial letters which will be convenient for our
mapping procedure.
Under crossings of external legs the parameters transform as
\begin{alignat}{3}
\pi_{12}:\quad z &\to 1 + y - z && && \\
\pi_{34}:\quad z &\to 1 + y - z,&\quad x &\to 1/(x y),&\quad m^2 &\to m^2 x^2 y\,.
\end{alignat}
In this parametrisation we factor out a normalisation of the form~\eqref{measure}
but with $\mm$ replaced by $m$.
We find the alphabet
\begin{align}\label{n12alphabet}
\{ l_1,\ldots,l_{17} \} = \{ & 
 x, 1 + x, y, 1 - y, z, 1 - z, -y + z, 1 + y - z, 1 + x y, 1 + x z, 
 x y + z,
 \notag\\ &
 1 + y + x y - z,
 1 + x + x y - x z,
 1 + y + 2 x y - z +  x^2 y z,
 \notag\\ &
 2 x y + x^2 y + x^2 y^2 + z - x^2 y z,
 1 + x + y + x y + x y^2 - z - x z - x y z,
 \notag\\ &
 1 + y + x y + y^2 + x y^2 - z - y z - x y z
 \}
\end{align}
at the level of the differential equations and also of the solutions through to
weight 4.
This alphabet is shorter than the previous one and can not be reduced further
by forming ratios.

We construct a new functional basis consisting of $\Li_{2,2}$ functions,
classical polylogarithms $\Li_n$ $(n=2,3,4)$ and logarithms, similar to
the approach taken in~\cite{Bonciani:2013ywa,Gehrmann:2014bfa}.
The $\Li_{2,2}$ function can be written in $\G$-function notation according to
\begin{equation}
\Li_{2,2}(x_1,x_2) = \G\left(0, \frac{1}{x_1}, 0, \frac{1}{x_1 x_2}; 1\right) \,.
\end{equation}
Following the algorithm of \cite{Duhr:2011zq} we generate functional arguments
which are rational functions of $x$, $y$, $z$ and do not lead to new spurious
letters.
This implies that the arguments factorise into the letters of our
alphabet and their inverses.
We can therefore systematically scan for admissable $\Li_n$
arguments by constructing power products of letters, their inverses and $-1$.
A candidate argument $x_1$ is admissable exactly if $1-x_1$ factorises into the
letters of our alphabet and $-1$, since only in that case the introduction
of new letters is avoided.
Admissable arguments for $\Li_{2,2}$ functions are determined
by forming pairs of admissable $\Li_n$ arguments and requiring
for any such pair $(x_1,x_2)$ that the difference $x_1-x_2$ factorises
into the original letters and $-1$.

For the amplitude we need to evaluate also independent master
integrals with crossed kinematics, and we chose to implement these expressions
explicitly for evaluation time optimisation purposes.
We therefore directly construct a shared set of basis functions for uncrossed
and crossed variants of the master integrals and consequently close our
alphabet~\eqref{n12alphabet} under $\pi_{12}$ and $\pi_{34}$ by adding the
letters
\begin{equation}\label{n12cross} 
\{ l_{18},l_{19} \} =  \{ -x y + z + x z + x y z, -y + z + y z + x y z \}.
\end{equation}

We require all functions to be single valued and real over the entire physical
region of phase space.
As in \cite{Gehrmann:2014bfa}, we further tighten this constraint and select only
those $\Li_{2,2}(x_1,x_2)$ functions, for which their power series representation
\begin{align}
\Li_{2,2}(x_1,x_2) &= \sum_{j_1 = 1}^{\infty}\sum_{j_2 = 1}^{\infty}
  \frac{x_1^{j_1}}{(j_1+j_2)^2} \frac{(x_1 x_2)^{j_2}}{j_2^2}
\end{align}
is convergent, that is, their arguments fulfil
\begin{equation}
|x_1| < 1\,, \qquad |x_1 x_2| < 1\,.
\end{equation}

We wish to express our master
integrals in terms of these new functions and employ the coproduct-augmented
symbol calculus for that mapping, see~\cite{Duhr:2011zq,Duhr:2012fh,Duhr:2014woa}.
This step is computationally demanding due to the large number of possible
candidate functions.
Here, we profit from the reduction of the number of letters described above
which leads to a smaller set of candidate integrals for a given maximal
total degree of the arguments.
Furthermore, we employ a particularly efficient technique for the symbol
calculus, where we identify and match individual factors of products
directly at the level of the symbol~\cite{vonManteuffel:2013vja}.
In particular, this means we never need to construct products of polylogarithms
for our candidate functions which avoids a severe combinatorial blowup
for the linear algebra routines.
Using the coproduct we were able to express all master integrals in terms
of our new set of functions described above.
We stress that the success of this matching is not a priori obvious.
The explicit solutions for all of the master integrals are provided on \hepforge.

Concerning our primary motivation for changing the functional basis,
we observe that the new representation indeed allows for significantly
faster numerical evaluations.
For the numerical evaluation of the multiple polylogarithms we employ
the implementation~\cite{Vollinga:2004sn} in the {\tt GiNaC}
library~\cite{Bauer:2000cp}.
The exact evaluation time and the speedup due to the new functional basis
depend on the chosen point in phase space and on the required precision of the
result.
We tested some samples and observed speedup-factors between 8 and 85
when evaluating the 111 master integrals in the system of differential equations.
For the benchmark point of~\cite{Caola:2014iua},
the numerical evaluation with default precision takes 2250\;ms for the
``traditional'' G-functions (Section~\ref{sec:tradpolylogs})
and 120\;ms for the ``optimised'' functions (this Section)
on a single core of a standard desktop computer.

\section{UV renormalisation and IR subtraction}
\label{sec:helfin}
Let us go back to the calculation of the helicity amplitude coefficients $A_j$ 
(or equivalently the $E_j$).
In order to simplify the notation for what follows we pick one of the
form factors:
\begin{equation}
\Omega = A_j ~~(\text{or~} E_j)\,, \quad \text{for some~} j =1,\ldots,10~(9)\,, 
\end{equation}
in order to suppress the index $j$.
The following discussions applies to any of the chosen form factors in the same way.

We perform renormalisation of the UV divergences in the standard $\MS$ scheme which,
in massless QCD, amounts to simply replacing the bare coupling $\alpha_0$ with the renormalised one 
$\alpha_s = \alpha_s(\mu^2)$, where $\mu^2$ is the renormalisation scale.
Since in our case the tree-level amplitudes do not contain any power of $\alpha_s$
we require only the one-loop relation for the coupling
\begin{equation}
\alpha_0\, \mu_0^{2 \epsilon}\, S_\epsilon = \alpha_s\, \mu^{2 \epsilon}
\left [ 1 - \frac{\beta_0}{\epsilon} \left( \frac{\alpha_s}{2 \pi} \right)+ \mathcal{O}(\alpha_s^2)\right] 
\end{equation}
where
\begin{equation}
S_\epsilon = (4 \pi)^\epsilon \, \e^{-\epsilon \gamma}\,, 
\qquad \mbox{with the Euler-Mascheroni constant} \quad \gamma = 0.5772...\,,
\end{equation}
$\mu_0^2$ is the mass-parameter introduced in dimensional regularisation to maintain 
a dimensionless coupling in the bare QCD Lagrangian density, and finally $\beta_0$
is the first order of the QCD $\beta$-function
\begin{equation}
\label{beta0}
\beta_0 = \frac{11 \,C_A - 4 \,T_F\,N_f}{6}\,, \quad \mbox{with}\quad C_A = N\,, 
\quad C_F = \frac{N^2-1}{2\,N}\,, \quad
T_F = \frac{1}{2}\,.  
\end{equation}
We perform UV renormalisation at the scale $\mu^2 = s$, the invariant mass of
the vector boson pair. Values of the helicity coefficients at 
different renormalisation scales can be recovered by using the renormalisation group equation. 
Since at a given loop order $n$ the form factors are
defined with all powers of the strong coupling factored out,
the renormalised form factors $\Omega^{(n)}$
are expressed in terms of the un-renormalised ones $\Omega^{(n),\un}$ according to
\begin{align}
&\Omega^{(0)} = \Omega^{(0),\un}\,,\nonumber \\
&\Omega^{(1)} = S_\epsilon^{-1}\,\Omega^{(1),\un}\,,\nonumber \\
&\Omega^{(2)} = S_\epsilon^{-2}\,\Omega^{(2),\un} 
- \frac{\beta_0}{\epsilon}\, S_\epsilon^{-1}\,\Omega^{(1),\un}\,. \label{UVren}
\end{align}

After performing UV renormalisation, the amplitude contains residual IR singularities
which will be cancelled analytically by those occurring in radiative processes at
the same order. Catani was the first to show how to organise the IR-pole
structure up to two-loop in QCD~\cite{Catani:1998bh}. In subtracting the poles from the one- and
two-loop amplitudes we will follow a slightly modified scheme described in~\cite{Catani:2013tia}, which
is better suited for the $q_T$-subtraction formalism.
The two schemes are of course equivalent and we provide formulae to convert the
results between the two schemes in Appendix~\ref{sec:catani}.
We define the IR finite amplitudes at renormalisation scale $\mu^2$
in terms of the UV renormalised ones as follows
\begin{align}
&\Omega^{(1),\finite} = \Omega^{(1)} - I_1(\epsilon)\, \Omega^{(0)}\,,\nonumber \\
&\Omega^{(2),\finite} = \Omega^{(2)} - I_1(\epsilon)\, \Omega^{(1)} - I_2(\epsilon)\, \Omega^{(0)}\,, \label{IRqT}
\end{align}
with
\begin{equation}
I_1(\epsilon) = I_1^{soft}(\epsilon) + I_1^{coll}(\epsilon) \,,
\end{equation}
\begin{align}
I_1^{soft}(\epsilon) &= -\frac{\e^{\epsilon \gamma}}{\Gamma(1-\epsilon)} \left( \frac{\mu^2}{s}\right)^{\epsilon}\, 
\left( \frac{1}{\epsilon^2} + \frac{i \pi}{\epsilon} + \delta_{q_T}^{(0)} \right)\,C_F
\,,\qquad
I_1^{coll}(\epsilon) =- \frac{3}{2}C_F \frac{1}{\epsilon} \left( \frac{\mu^2}{s}\right)^{\epsilon}\, ,
\end{align}
\begin{equation}
I_2(\epsilon) = - \frac{1}{2} I_1(\epsilon)^2 
+ \frac{\beta_0}{\epsilon} \left[ I_1(2 \epsilon) - I_1(\epsilon) \right]
+ K\,I_1^{soft}(2\epsilon) + H_2(\epsilon)\,,
\end{equation}
\begin{equation}
H_2(\epsilon) = \frac{1}{4 \epsilon}  \left( \frac{\mu^2}{s}\right)^{2\,\epsilon}\left( \frac{\gamma_q^{(1)}}{4} + C_F\, d_1 + 
\epsilon \, C_F\, \delta_{q_T}^{(1)} \right)\,,
\end{equation}
and the constants are defined as
\begin{align}
&\delta_{q_T}^{(0)} = 0\,, \qquad 
K = \left( \frac{67}{18} - \frac{\pi^2}{6} \right) C_A - \frac{5}{9} N_F \,,\nonumber \\
&d_1 = \left( \frac{28}{27} - \frac{1}{3} \zeta_2 \right) N_F 
+ \left( -\frac{202}{27} + \frac{11}{6}\,\zeta_2 + 7 \,\zeta_3 \right)  C_A\,,\nonumber \\
&\delta_{q_T}^{(1)} = \frac{10}{3} \zeta_3 \,\beta_0 
              + \left( -\frac{1214}{81} + \frac{67}{18} \zeta_2 \right) C_A 
              + \left( \frac{164}{81} - \frac{5}{9} \zeta_2 \right) N_F\,, \nonumber \\
& \gamma_{q}^{(1)}      = \left( -3 + 24\,\zeta_2 - 48 \zeta_3 \right) C_F^2 
             + \left( -\frac{17}{3} - \frac{88}{3} \zeta_2 + 24\,\zeta_3 \right) \,C_F\,C_A
             + \left( \frac{2}{3} + \frac{16}{3} \zeta_2 \right) \, C_F\, N_F\,.         
\end{align}
Note that in these equations all imaginary parts are already explicit prior to expansion in $\epsilon$.
Setting $\mu^2=s$, we calculated the finite remainder of the $A_j$ for $\epsilon \to 0$ in the $q_T$-subtraction scheme.
We provide the explicit analytical results on our project page at \hepforge.
It is straight-forward to convert our finite results obtained in the $q_T$-scheme to Catani's original scheme, see Appendix~\ref{sec:catani}.

\section{Checks on the amplitudes}
\label{sec:checks}

We performed different checks on our amplitude, which we enumerate here.
\begin{enumerate}
\item First of all, we started off by computing the 10 form factors $A_j$ of~\eqref{tensorstr}
for the
different classes of diagrams $\C=A,B,C,D_1,D_2$, and we explicitly verified that,
according to Furry's theorem, the diagrams in classes $D_1$ and $D_2$
independently sum up to zero.

\item From the $A_j$ we computed the 9 form factors $E_j$ in~\eqref{MLLL} and~\eqref{MRLL},
and we verified that, both prior to as well as after subtraction of UV and IR poles,
all symmetry relations described in~\eqref{Ejx12},\eqref{Ejx34} and the corresponding ones
for the $A_j$, are identically satisfied.

\item We verified that the poles of the one-loop and two-loop amplitudes are correctly
reproduced by Catani's formula~\cite{Catani:1998bh}, which provides a strong check on the
calculation.

\item For the NNLO computation of on-shell $ZZ$ and $W^+W^-$
production~\cite{Cascioli:2014yka, Gehrmann:2014fva}
we performed a dedicated calculation, directly for the squared amplitude,
employing our equal-mass master integrals~\cite{Gehrmann:2013cxs}.
The tree and one-loop contributions have been found to agree
with the analytical results of~\cite{Mele:1990bq,Frixione:1993yp}
and with numerical samples obtained with \openloops~\cite{Cascioli:2011va}.
Starting from our general results for the amplitude in the off-shell
case, we re-derived the squared amplitudes for on-shell $ZZ$ and $WW$ production
as described in Appendix~\ref{sec:equalmass} and found full agreement
through to two-loops.

\item We performed a thorough comparison of our results with an earlier calculation
of the two-loop amplitudes for on-shell $W^+\,W^-$ production in the small-mass limit~\cite{Chachamis:2008yb}.
Starting from our results for the squared amplitude for $W^+\,W^-$ production
(see Appendix~\ref{sec:equalmass}),
we take the small-mass limit, namely $m_W^2/s \to 0$ for fixed $(t-m_W^2)/s$.
Adjusting for overall conventions we found agreement with
the results obtained in~\cite{Chachamis:2008yb} in all contributions, 
except for $F_i^{[C],(2)}(s,t)$ arising from the interference of two-loop diagrams in class $C$ with the tree-level diagram in class $A$.
From the discussion in~\cite{Chachamis:2008yb},
we could trace back this discrepancy to a different treatment of the vector-axial contributions in the fermionic loop in class $C$, 
resulting in a non-vanishing remainder even for zero-mass quarks.
Since this appears to be inconsistent with charge parity conservation,
we have good reasons to believe that the prescription used here
as well as in~\cite{Caola:2014iua} is the correct one
(see our discussion in Section~\ref{sec:calc}).
 
\item Finally, we have compared numerically results both for the individual form factors
$E_j$ and for the full amplitudes $\M_{LLL}$ and $\M_{RLL}$
at tree-level, one-loop and two-loop order, with reference~\cite{Caola:2014iua}.
For the numerical evaluations of the helicity amplitudes we employed the
package {\tt S@M}~\cite{Maitre:2007jq}.
We find full agreement with the results reported in~\cite{Caola:2014iua},
after a mistake in the calculation of one of the form factors was corrected in that reference.
\end{enumerate}

\section{Numerical code and results}
\label{sec:numerics} 
\begin{figure}
\includegraphics[width=\textwidth]{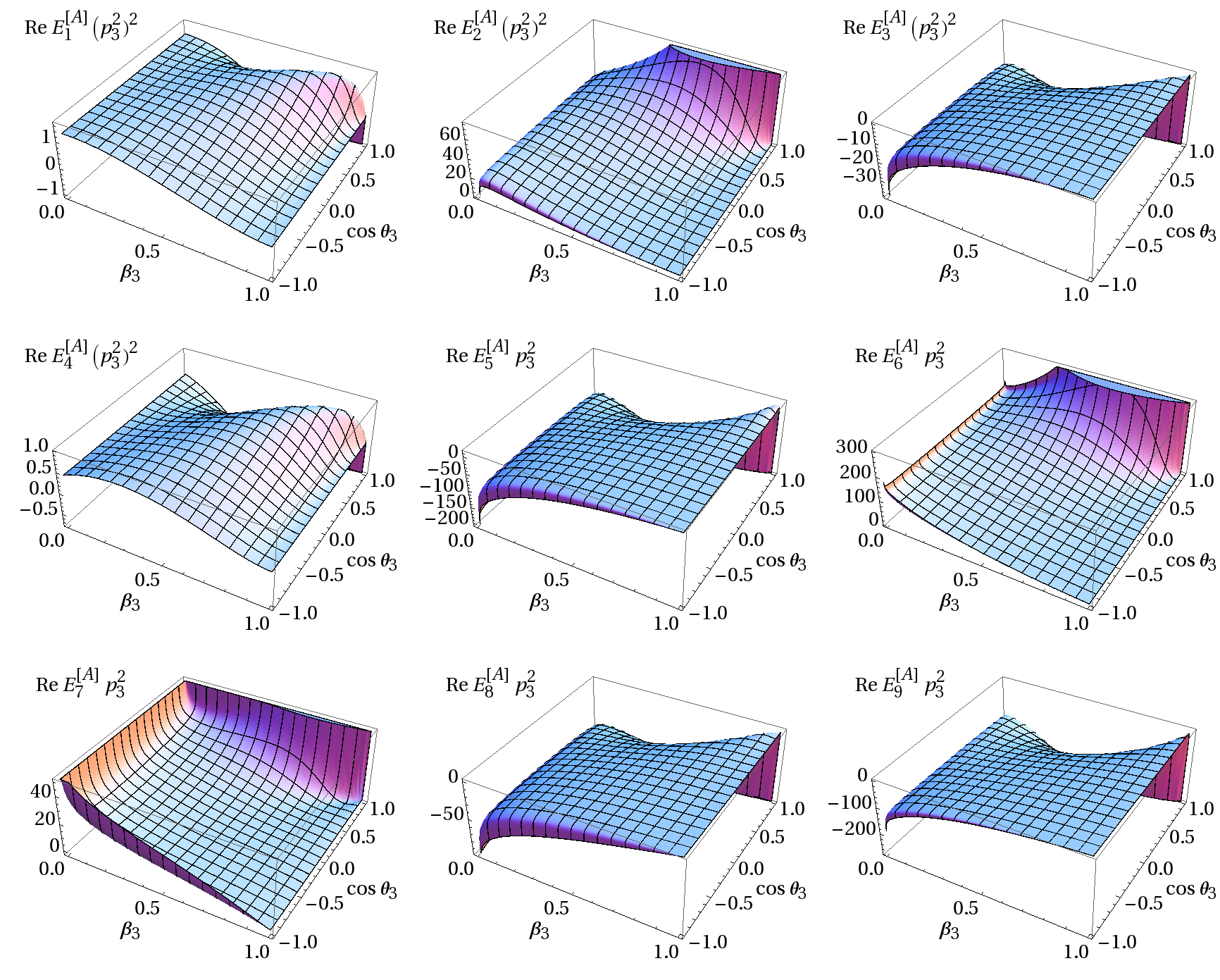}
\caption{\label{fig:ejclassa}
Real parts of the two-loop form factors $E_j^{[A],{(2)}}$ for the process $q\bar{q}'\to V_1 V_2$
in dependence of the relativistic velocity, $\beta_3$,
and the cosine of the scattering angle, $\cos\theta_3$, of the vector boson $V_1$.
The virtualities of the vector bosons are set to $p_4^2 = 2 p_3^2$.
}
\end{figure}
\begin{figure}
\includegraphics[width=\textwidth]{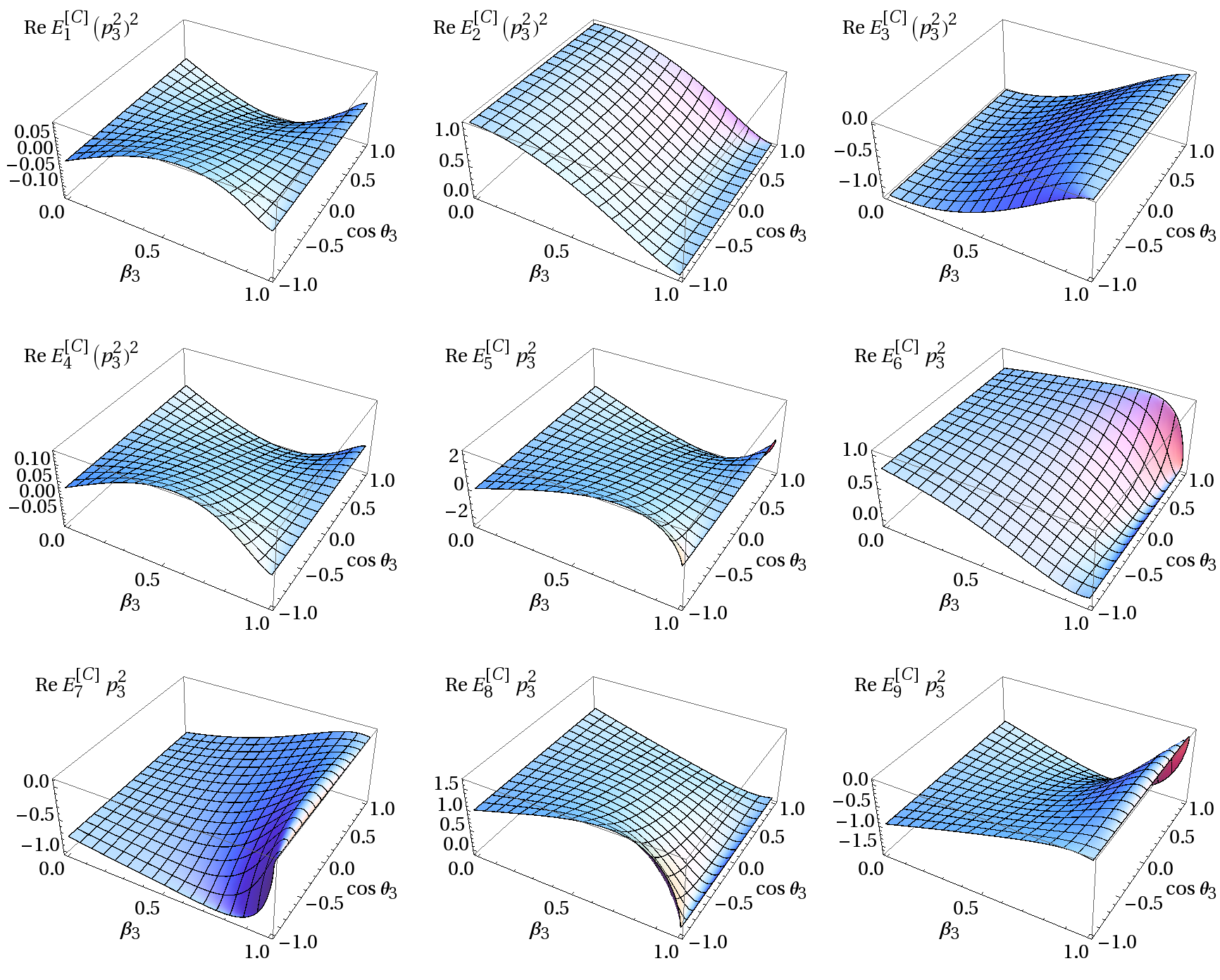}
\caption{\label{fig:ejclassc}
Real parts of the two-loop form factors $E_j^{[C],{(2)}}$ for the process $q\bar{q}'\to V_1 V_2$
in dependence of the relativistic velocity, $\beta_3$,
and the cosine of the scattering angle, $\cos\theta_3$, of the vector boson $V_1$.
The virtualities of the vector bosons are set to $p_4^2 = 2 p_3^2$.
}
\end{figure}
We provide a {\tt C++} code for the numerical evaluation of the 9 finite form factors $E_j$
for classes $A$, $B$ and $C$.
The implementation supports both, evaluation in the $q_T$-scheme and in Catani's original scheme.
Further, it also provides the (alternative) 10 form factors $A_j$.
The code is set up in form of a {\tt C++} library, which is supplemented by
a simple command line interface.

The code was optimised for speed and stability of the numerical evaluations, in particular, by
employing an appropriate functional basis for the multiple polylogarithms, see
Section~\ref{sec:optpolylogs}.
We employ {\tt C++} templates to support evaluations with three different data types:
double precision, quad precision and arbitrary precision using the {\tt CLN} library~\cite{cln}.
The multiple polylogarithms are evaluated via their implementation~\cite{Vollinga:2004sn}
in the {\tt GiNaC} library~\cite{Bauer:2000cp}, which also employs the {\tt CLN} arbitrary
precision capabilities.

For the benchmark point of~\cite{Caola:2014iua} no severe cancelations due to asymptotic kinematics
take place.
In this case our double precision implementation is accurate and
gives at least 10 significant digits for each of the $E_j$ at the two-loop level.
The evaluation of all $E_j$ incl.\ crossed variants, as needed for the physical amplitude,
takes 150\;ms on a single core of a standard desktop computer.
Close to the phase space boundaries or in the high energy region, numerical cancelations
lead to a significant loss of precision.
In order to detect and cure a possible instability, we compare the results obtained from evaluations with
different precision settings and adaptively increase the precision until the target precision is met.
We find the method to converge even in highly collinear configurations, where one needs to allow
for a significant increase in the evaluation time though.
Of course, also for unproblematic points in the bulk of the phase space, where
the double precision results are actually accurate enough, our precision
check requires additional run-time.
For the aforementioned benchmark point we find an increase in the evaluation
time to approximately 0.8\;s on a single core.

In Figures~\ref{fig:ejclassa} and \ref{fig:ejclassc} we show numerical results for the class $A$ and
class $C$ contributions to our 9 form factors $E_j$ at the two-loop level.
Note that these results were obtained with our {\tt C++} code and thus demonstrate
the high numerical reliability of our implementation.
We vary the relativistic velocity, $\beta_3 = \kappa/(s + p_3^2 - p_4^2)$,
and the cosine of the scattering angle, $\cos\theta_3 = (2 t + s - p_3^2 - p_4^2)/\kappa$,
of the vector boson $V$.
For the virtualities of the vector bosons we have set $p_4^2 = 2 p_3^2$.
All results are for $N_f=5$ and given in the $q_T$-scheme.
The class~$A$ contributions in Figure~\ref{fig:ejclassa} show pronounced
structures in the collinear regions (see \eqref{HelAmplTreeEA} for the corresponding tree level
coefficients).
In contrast, the class~$C$ contributions in Figure~\ref{fig:ejclassc} show no such features and are rather smooth functions
in the full $\beta_3$-$\cos\theta_3$ plane.

\section{Conclusions}
\label{sec:conc}
In this paper, we presented the derivation of the two-loop massless QCD 
corrections to the helicity amplitudes for massive 
vector boson pair production in quark-antiquark annihilation. 
The combination with leptonic decay currents allows to 
construct the two-loop QCD matrix elements relevant to four-lepton production. 
In this course, we computed all master integrals and optimised their representation
for numerical performance.
Our results obtained for the amplitudes provide a fully independent validation of
 a recent calculation~\cite{Caola:2014iua}.
We implemented our amplitudes in a {\tt C++} code for the fast
and stable numerical evaluation of the amplitudes, which we provide
together with our analytical results for public access at
\url{http://vvamp.hepforge.org}.
This opens up the path towards precision phenomenology in gauge boson pair
production and improvements of the
background predictions for Higgs boson studies 
and searches for physics beyond the Standard Model.

\section*{Acknowledgements}
We are grateful to K.~Melnikov and F.~Caola for their help in checking our results
against~\cite{Caola:2014iua}. LT wishes to thank K.~Melnikov
for a clarifying discussion on the use of 4-dimensional Schouten identities
for simplifications of spinor structures.
This research was supported in part by the Swiss National Science Foundation (SNF) 
under contract  200020-149517 and by 
the Research Executive Agency (REA) of the European Union under the Grant Agreement
 PITN--GA--2012--316704 ({\it HiggsTools}), and the ERC Advanced Grant {\it MC@NNLO} (340983).
We thank the \hepforge\ team for providing web space for our project.
The Feynman graphs in this article have been drawn with 
\jaxodraw~\cite{Binosi:2003yf,Vermaseren:1994je}.

\appendix

\section{Squared amplitudes for the on-shell production of vector-boson pairs}
\label{sec:equalmass}

In this Section we show how the general results described in this article can be used
to obtain the squared amplitude for the process $q \bar{q}' \to V_1 V_2$ summed over
spins and colours.
For the calculations of the NNLO QCD corrections to on-shell $ZZ$~\cite{Cascioli:2014yka}
and $W^+ W^-$ production~\cite{Gehrmann:2014fva} production, we directly
computed the squared amplitudes using a dedicated setup based on our solutions
for the equal-mass master integrals~\cite{Gehrmann:2013cxs,Gehrmann:2014bfa}.
We compared the results obtained in the two approaches and find full agreement.

We denote the squared amplitude as
\begin{equation}
 \langle \mathcal{M} | \mathcal{M} \rangle =\mathcal{T}(s,t,p_3^2,p_4^2) =  \sum_{pol, colour} 
 \left| S_{\mu \nu}(p_1,p_2,p_3) \epsilon_3^\mu(p_3) \epsilon_4^\nu(p_4)\right|^2\,,
\end{equation}
which of course can be perturbatively expanded in powers of $\alpha_s$ as

\begin{align}
 \mathcal{T}(s,t,p_3^2,p_4^2) &=(4 \pi \alpha)^2 \,
 \Bigg[ \mathcal{T}^{(0)}(s,t,p_3^2,p_4^2) + \left( \frac{\alpha_s}{2 \pi} \right)\mathcal{T}^{(1)}(s,t,p_3^2,p_4^2)\nonumber \\
 &+ \left( \frac{\alpha_s}{2 \pi} \right)^2 \mathcal{T}^{(2)}(s,t,p_3^2,p_4^2) + \mathcal{O}(\alpha_s^3) \Bigg]\,,
\end{align}
where we have

\begin{align}
& \mathcal{T}^{(0)}(s,t,p_3^2,p_4^2) = \langle \mathcal{M}^{(0)} | \mathcal{M}^{(0)} \rangle\,, \\
& \mathcal{T}^{(1)}(s,t,p_3^2,p_4^2) = 2 \Re \left( \langle \mathcal{M}^{(0)} | \mathcal{M}^{(1)} \rangle \right)\,, \\
& \mathcal{T}^{(2)}(s,t,p_3^2,p_4^2) = 2 \Re \left( \langle \mathcal{M}^{(0)} | \mathcal{M}^{(2)} \rangle \right)
+ \langle \mathcal{M}^{(1)} | \mathcal{M}^{(1)} \rangle\,.
 \end{align}

It is easy to write a general expression of $\langle \mathcal{M}^{(n)} | \mathcal{M}^{(m)} \rangle$ 
in terms of the coefficients $A_j^{(n)}$ and $A_j^{(m)}$
or, equivalently, in terms of the  $\tau_j^{(n)}$ and $\tau_j^{(m)}$, simply by
contracting the general decomposition~\eqref{tensorstr} with itself and summing over colours 
and external polarisations using~\eqref{polsum}. The result is quite involved
and not particularly illuminating and we decided not to include it here explicitly.
This general formula, in fact, is needed explicitly only in order to derive the $1$-loop$\times$$1$-loop
corrections $\langle \mathcal{M}^{(1)} | \mathcal{M}^{(1)} \rangle$, which can however also be
easily extracted from automated codes, and therefore we will not consider them here.
On the other hand, if we limit ourselves to considering the contraction of the generic 
$n$-loop amplitude with the tree-level, i.e. $m=0$, the results are much more compact.
In the following two sections we will discuss the two explicit cases of on-shell $ZZ$ and $WW$ production,
which were used for the calculations in~\cite{Cascioli:2014yka,Gehrmann:2014fva}.

\subsection{\texorpdfstring
{The two-loop corrections to $ZZ$ production}
{The two-loop corrections to ZZ production}
}
 In the case of $q \bar{q} \to ZZ$ the tree-level is given by the two diagrams belonging to classes $\C=A,B$. 
  As far as two-loop corrections are concerned, the classes of diagrams that can contribute to $ZZ$ production are
 $\C=A,B,C$, see Section~\ref{sec:calc}. By contracting the tree-level diagrams with the general amplitude~\eqref{tensorstr} 
 one easily finds
   \begin{equation}
      \langle \mathcal{M}^{(0)} | \mathcal{M}^{(n)}  \rangle_{ZZ} 
             =  \frac{N}{2}\left[ (L^Z_{qq})^4 + (R^Z_{qq})^4\right]\,\left( \frac{ 2\, \tau_8^{(ZZ,(n))} - \tau_{9}^{(ZZ,(n))} }{u}
             - \frac{ 2\, \tau_7^{(ZZ,(n))} - \tau_{10}^{(ZZ,(n))} }{t} 
        \right)\,,
       \end{equation}
      where $N$ is the number of colours, while $L_{qq}^Z$ and  $R_{qq}^Z$ are defined in~\eqref{ZLRcoupl}.
    Each of the $\tau_j^{(ZZ,(n))}$ can be obtained summing over the relevant classes of diagrams, re-weighted by appropriate coupling factors
       \begin{equation}
        \tau_j^{(ZZ,(n))} =  \tau_j^{[A],(n)} + \tau_j^{[B],(n)} + \widetilde{N}_{ZZ}\,\tau_j^{[C],(2)}\,,
       \end{equation}
      where the $\tau_j^{[\C]}$ components of the $\tau_j$ are defined by a decomposition
      completely analogous to that for the $A_j$ in \eqref{AdecompEW},
      \begin{equation}
      \widetilde{N}_{ZZ} = \frac{\left[ (L^Z_{qq})^2 + (R^Z_{qq})^2\right]}{\left[ (L^Z_{qq})^4 + (R^Z_{qq})^4\right]}\, N_{ZZ}
      \end{equation} 
      and $N_{ZZ}$ is defined in~\eqref{coupl1}\,.
       We have verified explicitly that as far at the tree-level and one-loop corrections are concerned, we 
       have full agreement with the results in~\cite{Mele:1990bq}.
   Similar but much more lengthy formulas can be derived for 
$\langle \mathcal{M}^{(1)} | \mathcal{M}^{(1)} \rangle_{ZZ}$, and we do not report them here for brevity.
   
\subsection{\texorpdfstring
{The two-loop corrections to $W^+ W^-$ production}
{The two-loop corrections to W+W- production}
}
 Let us consider now the case of $q_i \bar{q}_i \to W^+ W^-$, where the index $i$ labels the flavour of the 
 initial state quarks, $q_i=(u,d)$.
 At the tree-level, this process receives contributions from three diagrams, one in class $A$
 and the other two in class $F_V$, with $V = Z, \gamma$. Let us start from the tree-level and one-loop corrections, where
 only diagrams in classes $\C=A,F_V$ can contribute.
 Following the notation of~\cite{Frixione:1993yp}, we separate the 
 contributions to the squared amplitude into three different form factors
 \begin{equation}
 \langle \mathcal{M}^{(0)} | \mathcal{M}^{(0)}  \rangle_{i,WW}
  = N\left[ c_i^{tt}\, F_i^{(0)}(s,t) - c_i^{ts}\,J_i^{(0)}(s,t) + c_i^{ss} K_i^{(0)}(s,t)\right]\,, \label{decFrix0}
 \end{equation}
 \begin{equation}
 2 \Re\left( \langle \mathcal{M}^{(0)} | \mathcal{M}^{(1)}  \rangle_{i,WW} \right)
  = N\left[ c_i^{tt}\, F_i^{(1)}(s,t) - c_i^{ts}\,J_i^{(1)}(s,t) + c_i^{ss} K_i^{(1)}(s,t)\right] \,. \label{decFrix1}
 \end{equation}
$F_i^{(n)}$ contains the squared contribution of diagrams in class $\C=A$
(i.e. diagrams where the production of the $W^+ W^-$ pair is not mediated through a $\gamma$ or a $Z$ boson). 
$J_i^{(n)}$ encapsulates instead the interference of the $F_V$-type diagrams (i.e. those where
the $W^+ W^-$ pair is produced via a $\gamma$ or a $Z$ virtual boson) with diagrams in class $\C=A$.
Finally $K_i^{(n)}$ is given by the interference of the 
$F_V$-type diagrams with themselves. 
Again, following closely~\cite{Frixione:1993yp} we define then
\begin{align}
 &c_i^{tt} = \frac{1}{16\, \sin^4{\theta_w}}\,,\nonumber \\
  &c_i^{ts} = \frac{1}{4\,s\, \sin^2{\theta_w}} 
 \left( e_{q_i} - c_{ZW^+W^-}\,L^Z_{q_i q_i} \frac{s}{s-m_Z^2} \right)\,,\nonumber\\
 &c_i^{ss} = \frac{1}{s^2}\, 
 \left[ \left( e_{q_i} - \frac{c_{ZW^+W^-} (L^Z_{q_i q_i} + R^Z_{q_i q_i}) }{2} \frac{s}{s-m_Z^2} \right)^2 
 + \left( \frac{c_{ZW^+W^-}(L^Z_{q_i q_i} - R^Z_{q_i q_i})}{2} \frac{s}{s-m_Z^2} \right)^2 \right] \label{couplFrix}
\end{align}
where, as always, $e_{q_i}$ is the quark charge in units of $e$, with $e>0$, and the electroweak couplings
$L_{q_i q_i}^Z$, $R_{q_i q_i}^Z$ and $c_{ZW^+W^-}$ are defined in~\eqref{ZLRcoupl} and~\eqref{coupl2}.

At two loops the decomposition~\eqref{decFrix1} must be enlarged since also diagrams belonging to class $\C = C$
start contributing to the amplitude. We therefore write the two-loop contribution as follows
\begin{align}
 2 \Re\left( \langle \mathcal{M}^{(0)} | \mathcal{M}^{(2)}  \rangle_{i,WW} \right)
  = N&\Big[c_i^{tt}\, F_i^{(2)}(s,t) + c_i^{[C],tt}\, F_i^{[C],(2)}(s,t) \nonumber \\ &- c_i^{ts}\,J_i^{(2)}(s,t) - c_i^{[C],ts}\,J_i^{[C],(2)}(s,t)
  + c_i^{ss} K_i^{(2)}(s,t)\Big]\,, \label{decFrix2}
 \end{align}
 where we introduced the new couplings
 \begin{align}
 &c_i^{[C],tt} = \frac{1}{32\, \sin^4{\theta_w}} N_g\,,\nonumber \\
&c_i^{[C],ts} = \frac{1}{4\,s\, \sin^2{\theta_w}} 
 \left( e_{q_i} - \frac{c_{ZW^+W^-}\,\left( L^Z_{q_i q_i} + R^Z_{q_i q_i} \right)}{2} \frac{s}{s-m_Z^2} \right)N_g\,. \label{couplFrix2}
\end{align}
Here, the new form factors $F_i^{[C],(2)}(s,t)$ and $J_i^{[C],(2)}(s,t)$ contain the contribution from the 
two-loop diagrams in class $\C=C$. 
In deriving~\eqref{couplFrix2} we used the fact that for a fermion loop with an attached $W$-pair we have
\begin{equation}
N_{WW} = \frac{1}{2} \sum_{q\,q'} \left( L_{q q'}^W  L_{q' q}^W \right) = \frac{1}{4 \sin^2{\theta_w}} N_g\,,
\end{equation}
where $N_g=N_f/2$ 
is the number of generations of massless quarks running in the loop. Note that because of the flavour-change
induced by the $W^\pm$ bosons, we limit ourselves to consider at most $N_f =4$ massless quarks $(u,d,c,s)$,
i.e. two generations $N_g=2$.
Finally, the form factor $K_i^{(2)}(s,t)$
receives contributions only from one class of diagrams, $\C = F_V$.

At tree level we find that the different form factors can be obtained from
\begin{align}
 F_i^{(0)}(s,t) &= \left( \frac{2 \,\tau_{10}^{[A],(0)}  -4\,\tau_{7}^{[A],(0)}}{t} \right)\,, \\ 
 J_i^{(0)}(s,t) &=  4 \left( \tau_7^{[A],(0)}  + \,\tau_8^{[A],(0)} \right) 
                    - 2\left( \,\tau_{9}^{[A],(0)} + \,\tau_{10}^{[A],(0)} \right) \,, \\ 
 K_i^{(0)}(s,t) &=  2 \left( \tau_7^{[F],(0)}  + \,\tau_8^{[F],(0)} \right)
 - \left( \,\tau_{9}^{[F],(0)} + \,\tau_{10}^{[F],(0)} \right)    \,.
\end{align}
At one loop and two loops we find instead
\begin{align}
 F_i^{(n)}(s,t) &= 2 \Re \left( \frac{2 \,\tau_{10}^{[A],(n)}  -4\,\tau_{7}^{[A],(n)}}{t} \right)\,, \nonumber \\ 
 J_i^{(n)}(s,t) &=  2 \Re \left[ 2 \left( \tau_7^{[A],(n)}  + \,\tau_8^{[A],(n)} \right) 
                    - \left( \,\tau_{9}^{[A],(n)} + \,\tau_{10}^{[A],(n)} \right)  + \,\frac{1}{2}J_i^{(0)}(s,t) \F^{(n)}(s) \right]\,,
                    \nonumber \\ 
 K_i^{(n)}(s,t) &= 2 \Re \left( K_i^{(0)}(s,t) \, \F^{(n)}(s) \right) \,,
\end{align}
and the two new form factors read
\begin{align}
&F_i^{[C],(2)}(s,t) = 2 \Re \left( \frac{2\,\tau_{10}^{[C],(2)} - 4\,\tau_7^{[C],(2)}  }{t} \right)\,, \\
 &J_i^{[C],(2)}(s,t) =  2 \Re \left[ 2 \left( \tau_7^{[C],(2)}  + \,\tau_8^{[C],(2)} \right) 
 - \left( \,\tau_{9}^{[C],(2)} + \,\tau_{10}^{[C],(2)} \right) \right]
 \,,
\end{align}
where $\F^{(n)}(s)$ are the $n$-loop QCD corrections to the quark form factor.
We have verified that the tree-level and one-loop corrections, in the limit of equal
virtualities of the massive vector bosons, agree with~\cite{Frixione:1993yp}.

\section{Schouten identities for the amplitude}
\label{sec:schouten}

In this Appendix, we show how to reduce the number of independent form factors entering
our helicity amplitudes by exploiting the 4-dimensionality of external states via Schouten identities.
We document here a general way to derive such Schouten identities for the $\M_{RLL}$ case. 
The $LLL$ case proceeds in exactly the same way.

We start off by fixing the helicities for a right-handed incoming quark current in
the spinor helicity notation and we get
\begin{align*}
S^{\mu \nu}_R(p_1^-, p_2^+,p_3) &= 
        [ 2\, \p_3\,1 \rangle \, \left( A_1\,p_1^\mu\, p_1^\nu
       + A_2\, p_1^\mu p_2^\nu + A_3 \, p_1^\nu p_2^\mu
       + A_4\,\,p_2^\mu\, p_2^\nu 
\right) \nonumber \\
       &+  [2\, \gamma^\mu \, 1 \rangle \left( A_5\, p_1^\nu + A_6 p_2^\nu \right)
        +  [2\, \gamma^\nu \, 1 \rangle \left( A_7\, p_1^\mu + A_8 p_2^\mu \right)\nonumber \\
       &+  A_9\, [2\, \gamma^\nu \p_3 \gamma^\mu \, 1 \rangle 
          + A_{10} \, [2\, \gamma^\mu \p_3 \gamma^\nu \, 1 \rangle \,.
\end{align*}
As a first step we notice that we can collect  $[ 2\, \p_3\,1 \rangle$ as an overall factor: 
\begin{align*}
[2\, \p_3\,1 \rangle [ 1\, \p_3\,2 \rangle  = {\rm Tr}\left[ \p_2\, \p_3\, \p_1\, \p_3\, 
\frac{1+\gamma_5}{2} \right]
= t\,u - p_3^2\, p_4^2\,.
\end{align*}
Multiplying and dividing by this allows to write the partonic amplitude as
\begin{align}
S^{\mu \nu}_R(p_1^-, p_2^+,p_3) &= 
        [ 2\, \p_3\,1 \rangle \, \Big\{ \left( A_1\,p_1^\mu\, p_1^\nu
       + A_2\, p_1^\mu p_2^\nu + A_3 \, p_1^\nu p_2^\mu
       + A_4\,\,p_2^\mu\, p_2^\nu 
\right) \nonumber \\
       &+  \frac{[1\, \p_3\,2 \rangle\, [2\, \gamma^\mu \, 1 \rangle }{t\,u-p_3^2 p_4^2} 
              \left( A_5\, p_1^\nu + A_6 p_2^\nu \right)
        +  \frac{[1\, \p_3\,2 \rangle\, [2\, \gamma^\nu \, 1 \rangle }{t\,u-p_3^2 p_4^2} 
        \left( A_7\, p_1^\mu + A_8 p_2^\mu \right)\nonumber \\
       &+ \frac{ A_9}{t\,u-p_3^2 p_4^2}\, [1\, \p_3\,2 \rangle\, [2\, \gamma^\nu \p_3 \gamma^\mu \, 1 \rangle 
          + \frac{A_{10}}{t\,u-p_3^2 p_4^2} \, [1\, \p_3\,2 \rangle \, [2\, \gamma^\mu \p_3 \gamma^\nu \, 1 \rangle\, \Big\},
          \label{Shelfix}
\end{align}
such that every spinor structure is a trace. We can then perform the traces recalling that
the transversality of the leptonic decay currents allows to discard contributions
proportional to $p_3^\mu$ or $p_4^\nu$.
In this way we get
\begin{align}
[1\, \p_3\,2 \rangle\, [2\, \gamma^\mu \, 1 \rangle\, &= 
 2\,\epsilon^{p_1,p_3,p_2,\mu}\, - (u-p_3^2) p_1^\mu  - (t-p_3^2)  p_2^\mu \label{trace1}
\end{align}
and 
\begin{align}
[1\, \p_3\,2 \rangle \, [2\, \gamma^\mu \p_3 \gamma^\nu \, 1 \rangle &=
2\,(u-p_3^2)\,\epsilon^{p_1,p_3,\mu,\nu} + 2\, p_3^2\, \epsilon^{p_1,p_2,\mu,\nu}
- (t\,u -p_3^2 p_4^2) g^{\mu \nu} \nonumber \\
& - 2\,u\,p_1^\mu p_2^\nu + 2\,p_3^2\, p_1^\nu p_2^\mu - 2\,(u-p_3^2) \,p_1^\mu p_1^\nu \label{trace2}
\end{align}
\begin{align}
[1\, \p_3\,2 \rangle \, [2\, \gamma^\nu \p_3 \gamma^\mu \, 1 \rangle &=
- 2\,(u-p_3^2)\,\epsilon^{p_1,p_3,\mu,\nu} - 2\, p_3^2\, \epsilon^{p_1,p_2,\mu,\nu}
+ 4\,\epsilon^{p_1,p_3,p_2,\mu} \left( p_1^\nu + p_2^\nu \right) \nonumber \\
&- (t\,u -p_3^2 p_4^2) g^{\mu \nu} 
 - 2\,t\,p_1^\nu p_2^\mu + 2\,p_3^2\, p_1^\mu p_2^\nu - 2\,(t-p_3^2) \,p_2^\mu p_2^\nu\,, \label{trace3}
\end{align}
where we introduced the Levi-Civita $\epsilon$ tensor, with the following notation
$$\epsilon^{p,q,r,s} = \epsilon^{\mu,\nu,\rho,\sigma} p_\mu q_\nu r_\rho s_\sigma\,.$$
Moreover, note that the asymmetry between~\eqref{trace2} and~\eqref{trace3} is due to the transversality
condition which effectively replaces $p_3^\mu \to 0$ and $p_3^\nu \to p_1^\nu + p_2^\nu$\,.

Using~\eqref{trace1},\eqref{trace2} and~\eqref{trace3} we see that all $10$ spinor structures
can be written in terms of the following 11 structures:
$$
g^{\mu \nu}\,, \qquad p_1^\mu p_1^\nu\,, \quad p_1^\mu p_2^\nu\,, \quad p_2^\mu p_1^\nu\,, \quad p_2^\mu p_2^\nu\,, $$
$$\epsilon^{p_1,p_3,p_2,\mu}\,p_1^\nu\,, \quad \epsilon^{p_1,p_3,p_2,\mu}\,p_2^\nu \,, \quad
\epsilon^{p_1,p_3,p_2,\nu}\,p_1^\mu\,, \quad \epsilon^{p_1,p_3,p_2,\nu}\,p_2^\mu
$$
$$
\epsilon^{p_1,p_3,\mu,\nu}\,, \quad \epsilon^{p_1,p_2,\mu,\nu}\,.
$$

This does not appear to be any improvement with respect to the 10 structured we had before.
It is nevertheless very easy to show that $2$ out of these $11$ structures can indeed be expressed as 
linear combinations of the remaining $9$ by means of an anti-symmetrisation of the 
$\epsilon^{\mu \nu \rho \sigma}$ tensors.

In order to see how this works in practice, we start off by considering 
$\epsilon^{p_1,p_3,\mu,\nu}\, p_2 \cdot p_1$. By anti-symmetrising
 $\epsilon^{\mu,\nu,\rho,\sigma}p_2^\tau$ in $4$ dimensions one easily finds
\begin{align}
\epsilon^{p_1,p_3,\mu,\nu}\, p_2 \cdot p_1 &= - \epsilon^{p_3,\mu,\nu,p_2}\, p_1 \cdot p_1
                                                                          -  \epsilon^{\mu,\nu,p_2,p_1}\, p_3 \cdot p_1
                                                                          -  \epsilon^{\nu,p_2,p_1,p_3}\,  p_1^\mu
                                                                          - \epsilon^{p_2,p_1,p_3,\mu}\,  p_1^\nu 
\end{align}
which implies that $\epsilon^{p_1,p_3,\mu,\nu}$ can be eliminated by
\begin{align}                                                                          
 \epsilon^{p_1,p_3,\mu,\nu}      &= \frac{2}{s}\, \left( \frac{p_3^2-t}{2}\epsilon^{p_1,p_2,\mu,\nu}\, 
                                                                          +  \epsilon^{p_1,p_3,p_1,\nu}\,  p_1^\mu
                                                                          - \epsilon^{p_1,p_3,p_2,\mu}\,  p_1^\nu \right)\,, \label{epsrel1}
\end{align}
leaving us again with $10$ structures. One more anti-symmetrisation can be used, namely
consider $\epsilon^{p_1,p_2,\mu,\nu}\, p_3 \cdot r $, where the momentum $r^\mu$ is 
defined as
$$r^\mu = \left(\frac{u-p_3^2}{s}\right) p_1^\mu + \left(\frac{t-p_3^2}{s}\right) p_2^\mu + 	p_3^\mu\,,$$
such that $r \cdot p_1 = 0\,,\; r \cdot p_2 = 0$. Proceeding as before we find
\begin{align}
\epsilon^{p_1,p_2,\mu,\nu} &=  \frac{ \epsilon^{p_1,p_3,p_2,\mu} }{t\,u - p_3^2 p_4^2}  
\left[ (u - p_4^2) \,p_2^\nu
       +(t - p_4^2) \,p_1^\nu
         \right] 
         + \frac{ \epsilon^{p_1,p_3,p_2,\nu} }{t\,u - p_3^2 p_4^2}          
\left[      (u - p_3^2)\,p_1^\mu
          + (t - p_3^2)\,p_2^\mu
          \right] \label{epsrel2}\,.
\end{align}

It becomes clear that using these two relations we can eliminate completely $\epsilon^{p_1,p_2,\mu,\nu}$
and $\epsilon^{p_1,p_3,\mu,\nu}$ in favour of the remaining $9$ structures. In particular these relations can
be rephrased in terms of the original spinors in~\eqref{Shelfix} giving two Schouten identities for the spinor lines:
\begin{align}
[1\, \p_3\,2 \rangle \, [2\, \gamma^\mu \p_3 \gamma^\nu \, 1 \rangle &= 
(t \, u - p_3^2 p_4^2) \left[ \frac{2}{s} \left( p_1^\mu p_2^\nu - p_1^\nu p_2^\mu \right)
- g^{\mu \nu} \right] \nonumber \\
& +\frac{1}{s} \left[ (u-p_3^2) \, p_1^\mu - (t-p_3^2)\,p_2^\mu \right]
[1\, \p_3\,2 \rangle \, [2\, \gamma^\nu \, 1 \rangle \nonumber \\
& - \frac{1}{s} \left[ (u-s-p_3^2)\, p_1^\nu + (u-p_4^2)\,p_2^\nu \right] \, 
[1\, \p_3\,2 \rangle \, [2\, \gamma^\mu \, 1 \rangle\,, \label{Schouten1}
\end{align}

\begin{align}
[1\, \p_3\,2 \rangle \, [2\, \gamma^\nu \p_3 \gamma^\mu \, 1 \rangle &= 
(t \, u - p_3^2 p_4^2) \left[ \frac{2}{s} \left( p_1^\nu p_2^\mu - p_1^\mu p_2^\nu \right)
- g^{\mu \nu} \right] \nonumber \\
& +\frac{1}{s} \left[ (t-p_3^2) \, p_2^\mu - (u-p_3^2)\,p_1^\mu \right]
[1\, \p_3\,2 \rangle \, [2\, \gamma^\nu \, 1 \rangle \nonumber \\
& - \frac{1}{s} \left[ (t-p_4^2)\,p_1^\nu + (t-s-p_3^2)\, p_2^\nu \right] \, 
[1\, \p_3\,2 \rangle \, [2\, \gamma^\mu \, 1 \rangle\,. \label{Schouten2}
\end{align}

The corresponding relations for the spinors of the left-handed partonic currents
can be found by simply permuting $p_1 \leftrightarrow p_2$.
Using~\eqref{Schouten1},\eqref{Schouten2}, and the corresponding ones for 
the left-handed partonic current, we eliminate $2$ of the structures
in~\eqref{Shelfix} in favour of $g^{\mu \nu}$, plus the remaining $8$ structures,
and then proceed by contracting with the left-handed leptonic
decay currents~\eqref{LepCurrL}. As a result one easily arrives at
formulae~\eqref{MLLL} and \eqref{MRLL}.

\section{Conversion to Catani's original IR subtraction scheme}
\label{sec:catani}
In Section~\ref{sec:helfin} we derived the finite remainder of the one- and two-loop helicity amplitude
coefficients $\Omega$ in a subtraction scheme which is particularly well-suited for 
$q_T$ subtraction~\cite{Catani:2013tia}.
In this Appendix we show how these results can be converted to Catani's original scheme~\cite{Catani:1998bh}.
Starting from the UV-renormalised coefficients defined in~\eqref{UVren} 
at renormalisation scale $\mu^2$, 
we write the finite remainders in Catani's scheme as
\begin{align}
\Omega^{(1),\finite}_{\text{Catani}} &= \Omega^{(1)} - I_1^{\text{C}}(\epsilon) \,\Omega^{(0)}\,,\nonumber \\
\Omega^{(2),\finite}_{\text{Catani}} &= \Omega^{(2)} - I_1^{\text{C}}(\epsilon) \,\Omega^{(1)}  
- I_2^{\text{C}}(\epsilon)  \,\Omega^{(0)}\,, \label{IRCatani}
\end{align}
where Catani's subtraction operators are defined as
\begin{align}
I_1^{\text{C}}(\epsilon)  &=  -C_F \frac{e^{\epsilon \gamma}}{\Gamma(1-\epsilon)} 
\left(\frac{1}{\epsilon^2} + \frac{3}{2\epsilon} \right) 
\left(-\frac{\mu^2}{s}\right)^{\epsilon}\nonumber \\
I_2^{\text{C}}(\epsilon)  &= -\frac{1}{2} I_1^C(\epsilon) \left(I_1^C(\epsilon)
+\frac{2\beta_0}{\epsilon} \right)
+\frac{e^{-\epsilon \gamma} \Gamma(1-2\epsilon)} {\Gamma(1-\epsilon)}
\left(\frac{\beta_0}{\epsilon}+ K \right) I_1^C(2 \epsilon) + H^{(2)}(\epsilon) 
\end{align}
with
\begin{align}
K&=\left(\frac{67}{18} -\frac{\pi^2}{6}\right) C_A -\frac{10}{9} T_F N_f\,,
\end{align}
and since a $q \bar q$ pair is the only coloured state we have 
\begin{align}
H^{(2)}(\epsilon) &= \frac{e^{\epsilon \gamma}}{4 \epsilon \Gamma(1-\epsilon)} 
\left(-\frac{\mu^2}{s}\right)^{2\epsilon} \notag\\
&\times
2 C_F
\left[
\left(\frac{\pi^2}{2}-6\,\zeta_3 
-\frac{3}{8}\right) C_F + \left(\frac{13}{2}\zeta_3 
+ \frac{245}{216}-\frac{23}{48} \pi^2 \right) C_A 
+  \left(\frac{\pi^2}{12} -\frac{25}{54}\right)  T_F N_f 
\right]\, .
\end{align}
In this article, we present our results for $\mu^2=s$.
Note that upon expansion in $\epsilon$ both $I_1^C(\epsilon)$ and $I_2^C(\epsilon)$
generate imaginary parts whose sign is fixed by the prescription $s \to s + i \,0^+$\,.

By comparing~\eqref{IRqT} with~\eqref{IRCatani} one can show that 
the $\epsilon^0$ parts of the finite, complex form factors
in Catani's original scheme~\cite{Catani:1998bh},
can be obtained from those in the 
$q_T$-scheme~\cite{Catani:2013tia} according to
\begin{align}\label{qt2catani}
\Omega^{(1),\finite}_{\text{Catani}} &= \Omega^{(1),\finite}_{q_T} + \Delta I_1 \,\Omega^{(0),\finite}_{q_T},\nonumber\\
\Omega^{(2),\finite}_{\text{Catani}} &= \Omega^{(2),\finite}_{q_T} + \Delta I_1 \, \Omega^{(1),\finite}_{q_T} + \Delta I_2 \,\Omega^{(0),\finite}_{q_T},
\end{align}
with the finite scheme conversion coefficients given by
\begin{align}\label{qt2catanicoeff}
 \Delta I_1 &= C_F \left( -\frac{1}{2} \pi^2 + i \pi \frac{3}{2} \right),\\
 \Delta I_2 &=
    C_A C_F \left( - \frac{607}{162}-\frac{1181}{432} \pi^2
            +\frac{187}{72} \zeta_3+\frac{7}{96} \pi^4
      +i \pi \left(\frac{961}{216}+\frac{11}{72} \pi^2-\frac{13}{2} \zeta_3\right) \right) \nonumber\\ &\quad
    +C_F^2 \left( -\frac{9}{8} \pi^2 + \frac{1}{8} \pi^4
      + i \pi \left(\frac{3}{8}-\frac{5}{4} \pi^2 + 6 \zeta_3\right) \right) \nonumber\\ &\quad
    + N_f C_F \left( \frac{41}{81}+\frac{97}{216} \pi^2-\frac{17}{36} \zeta_3
           + i \pi \left(- \frac{65}{108} - \frac{1}{36} \pi^2 \right)\right),
\end{align}
where we have set $\mu^2 = s$ to match the convention for our final results.
Notice that, in order to obtain the finite remainders of the two-loop amplitudes in the two different schemes,
only the finite pieces of the latter are required, and in particular the
$\mathcal{O}(\epsilon)$ terms of the one-loop amplitudes are not needed, as expected.
Note, moreover, that the conversion coefficients are complex, due to the fact that the 
original formulation of IR subtraction~\cite{Catani:1998bh} factored out a phase for time-like pairs of partons from both the 
collinear and soft contributions, while in the 
$q_T$-scheme~\cite{Catani:2013tia} this phase factor is associated only with the soft contributions,
in line with the structure of IR factorisation~\cite{Gardi:2009qi,Becher:2009qa} at higher loop order.

\bibliographystyle{JHEPvtag}   
\bibliography{Biblio}     

\end{document}